\documentclass[journal=ancac3,manuscript=article]{achemso}

\usepackage[version=3]{mhchem} 
\usepackage{xcolor,caption,amssymb}

\author{Athulya K. Muraleedharan}
\affiliation[LuMIn]
{Université Paris-Saclay, ENS Paris-Saclay, CNRS, CentraleSupélec, LuMIn, 91190 Gif-sur-Yvette, France}
\author{Kevin Co}
\author{Maxime Vallet}
\author{Abdelali Zaki}
\author{Fabienne Karolak}
\author{Christine Bogicevic}
\affiliation[SPMS]
{Université Paris-Saclay, CentraleSupélec, CNRS, Laboratoire SPMS, 91190 Gif-sur-Yvette, France}

\author{Karen Perronet}
\affiliation[LuMIn]
{Université Paris-Saclay, ENS Paris-Saclay, CNRS, CentraleSupélec, LuMIn, 91190 Gif-sur-Yvette, France}

\author{Brahim Dkhil}
\affiliation[SPMS]
{Université Paris-Saclay, CentraleSupélec, CNRS, Laboratoire SPMS, 91190 Gif-sur-Yvette, France}
\author{Charles Paillard}
\affiliation[SPMS]
{Université Paris-Saclay, CentraleSupélec, CNRS, Laboratoire SPMS, 91190 Gif-sur-Yvette, France}
\alsoaffiliation[UA]
{Smart Ferroic Materials, Institute for Nanoscience \& Engineering and Department of Physics, University of Arkansas, Fayetteville 72701 Arkansas, United States}

\author{Céline Fiorini-Debuisschert}
\affiliation[SPEC]
{Université Paris-Saclay, CEA, CNRS, SPEC, 91191 Gif-sur-Yvette, France}

\author{Fran\c cois Treussart}
\affiliation[LuMIn]
{Université Paris-Saclay, ENS Paris-Saclay, CNRS, CentraleSupélec, LuMIn, 91190 Gif-sur-Yvette, France}
\email{francois.treussart@ens-paris-saclay.fr}

\title[BTO NC ferroelectric texture]
  {Ferroelectric texture of individual barium titanate nanocrystals}


\begin{document}

\begin{tocentry}

\includegraphics{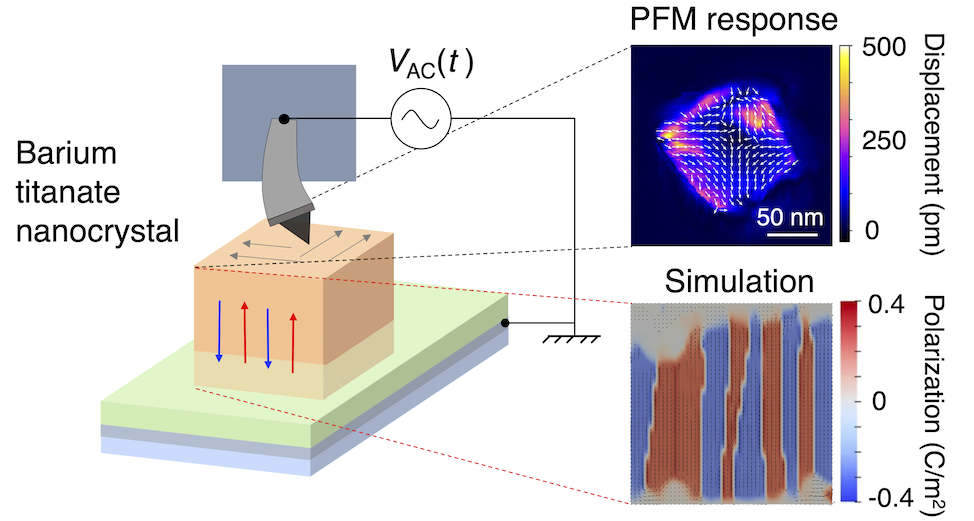}
Left: single barium titanate nanocrystal (its crystallographic axis $Z_{\rm c}$ being vertical) embedded in a conductive polymer layer (green), and with a piezoforce response microscopy (PFM) tip in contact with its top surface. Right, top: Mixed lateral PFM scan (in response to AC voltage modulation $V_{\rm AC}(t)$) displaying the displacement amplitude (color scale) and direction (white arrows). Right, bottom: phase field simulation of the equilibrium vertical component $P_{Z_{\rm c}}$ of the spontaneous polarization at zero external electric field, highlighting the core composed of 180° up and down domains. Note that $P_{Z_{\rm c}}$ vanishes on the top and bottom surfaces where the polarization in only in plane, leading to 90° domains in these facets.

\end{tocentry}

\begin{abstract}
  Ferroelectric materials display exotic polarization textures at the nanoscale that could be used to improve the energetic efficiency of electronic components. The vast majority of studies were conducted in two dimensions on thin films, that can be further nanostructured, but very few studies address the situation of individual isolated nanocrystals synthesized in solution, while such structures could have other field of applications. In this work, we experimentally and theoretically studied the polarization texture of ferroelectric barium titanate (BaTiO$_3$, BTO) nanocrystals (NC) attached to a conductive substrate and surrounded by air. We synthesized NC of well defined quasi-cubic shape and 160~nm average size, that conserve the tetragonal structure of BTO at room temperature. 
  
  We then investigated the inverse piezoelectric properties of such pristine individual NC by vector piezoresponse force microscopy (PFM), taking particular care of suppressing electrostatic artifacts. In all the NC studied, we could not detect any vertical PFM signal, and the maps of the lateral response all displayed larger displacement amplitude on the edges with deformations converging toward the center. Using field-phase simulations dedicated to ferroelectric nanostructures, we were able to predict the equilibrium polarization texture. These simulations revealed that the NC core is composed of 180° up and down domains defining the polar axis, that rotate by 90° in the two facets orthogonal to this axis, eventually lying within these planes forming a layer of about 10~nm thickness mainly composed of 180° domains along an edge. From this polarization distribution we predicted the lateral PFM response, that revealed to be in very good qualitative agreement with the experimental observations. This work positions PFM as a relevant tool to evaluate the potential of complex ferroelectric nanostructures to be used as sensors.

\end{abstract}

\section{Introduction}
Ferroelectric materials gather a set of physical properties, including a large dielectric permittivity, a piezoelectric response, and ferroelectric hysteresis that make them vital for several industries~\cite{Mikolajick2021}. Recently, progress in the synthesis and growth methods of ferroelectrics has allowed exotic effects to be discovered at low dimensions. For instance, quasiparticles such as (anti)vortices or polar skyrmions have been widely predicted~\cite{Naumov2004, Louis2012,Prosandeev2008a,Nahas2015,Nahas2017a,Goncalves2019} and subsequently observed~\cite{Balke2011,Yadav2016,Das2019,Abid2021,Tan2021} due to electrostatic depolarization in two-dimensional structures such as superlattices or thin films, or in one-dimensional nanowires embedded in a dielectric matrix. These topological defects are often endowed with rich properties, such as negative dielectric capacitance~\cite{Das2021}, an effect envisioned to reduce Field Effect Transistor energy consumption. The polarization texture in 0-dimensional ferroelectric crystals, that is, ferroelectric nanodots and nanocrystals (NC) have been less studied; yet recent (mainly simulation) studies have reported exotic polarization textures in NC as well~\cite{Li2017,Mangeri2017,Lukyanchuk2020,Co2021}. 

Therefore, in this work we set out to understand the polarization texture of ferroelectric BaTiO\textsubscript{3} (BTO) NC by an original combination of experiments and phase field modeling~\cite{Mangeri2017}. BTO is a prototypical ferroelectric material with a paraelectric phase above 120$^{\circ}$C in bulk crystal. Below that transition temperature, and down till about 5$^{\circ}$C, a polar order and a tetragonal deformation of the unit cell are established. This is the room temperature ferroelectric phase. Below 5$^{\circ}$C, a succession of ferroelectric orthorhombic and rhombohedral phases appears~\cite{Kwei.1993}. The synthesis of BTO nanocrystals with well-controlled morphology, size, and properties is well established~\cite{Bogicevic.2015,Jiang.2019}. 
Compared to bulk, the larger surface-to-volume ratio of BTO NC leads to starkly different properties, particularly ferroelectric ones, which strongly depend on boundary conditions. For instance, the phase transition from the para- to the ferro-electric state is more diffuse in BTO NC~\cite{Smith.2008}. Furthermore, the small dimensions offered in BTO NC, combined with the presence of electrical polarization, offer competitive advantages for charge separation in photovoltaic or photocatalytic applications~\cite{Cui2013,Paillard2016,Li2020a,Hao2021,Abbasi.2022,Neige.2023, Assavachin.2023}. However, there is little to no report on the nanoscale inhomogeneity of the polarization distributions, which may be critical for applications such as photocatalysis. 

A frequently used technique of characterization at the nanoscale is the high-angle annular dark-field (HAADF) imaging mode of scanning transmission electron microscopy (STEM), harnessing its high sensitivity to ion nuclei displacements. A polarization structure was observed by HAADF-STEM within 200~nm sized BTO free-standing cuboids milled in a bulk crystal using a focused ion beam. The observations were interpreted as the results of quadrant structures of 90$^\circ$ polarization domains~\cite{Schilling.2009}. As the nanocuboids were milled in bulk, one cannot exclude that the domain structure is partly induced by the fabrication process, as recently reported~\cite{Denneulin.2022}. Other electron beam-based techniques, capable of sensing the electric field of the sample via phase accumulation and its subsequent retrieval, are used to probe the local order in nanoferroelectrics~\cite{Campanini.2019}. In particular, Polking \textit{et al.} harnessed off-axis electron holography to prove that a 10~nm-sized BTO nanocube retains ferroelectric properties, as evidenced by the presence of an electric field around the NC only below Curie temperature~\cite{Polking.2012}. Moreover, taking advantage of the atomic resolution provided by their aberration-corrected STEM, they were able to reconstruct the atomic displacements in the 10~nm nanocube and demonstrated that it is made of a single ferroelectric domain.
Using Bragg coherent X-ray diffraction imaging of single BTO nanoparticles of larger size (160~nm) embedded in a non-ferroelectric polymer matrix, Karpov \textit{et al.}\cite{Karpov.2017} extracted much more complex polarization field (than in 10~nm-sized BTO nanocube) showing that it arranges in a flux-closure manner (forming a vortex), that can be consistently reproduced by phase field simulations.
Alternatively to these non-contact methods based on electron or X-ray beams, near field microscopy in contact mode has been used. For example, the ferroelectric properties of an array of 60~nm-sized dots nanostructured in a thin film of bismuth iron oxide grown on a conductive substrate have been explored by Piezoresponse Force Microscopy (PFM)~\cite{Li2017}, revealing various polarization textures in individual dots. 

Here, we combine PFM and phase field simulations to study the polarization texture of isolated BTO nanocubes in their native ferroelectric state, as produced by solvothermal synthesis.
We first present NC synthesis and structural characterization, then electromechanical   mapping on individual NC using PFM, and finally phase field simulations of such displacement fields. We show that the simulations are in very good qualitative agreement with the PFM data, enlarging the domain of application of PFM, mostly used for thin film characterization, to complex nanostructures, for which phase field simulations can be demanding.

\section{Results and discussion}
\subsection{Barium titanate nanocrystals of quasi-cubic shape}
\begin{figure}[h!]
    \centering
    \includegraphics[width=\textwidth]{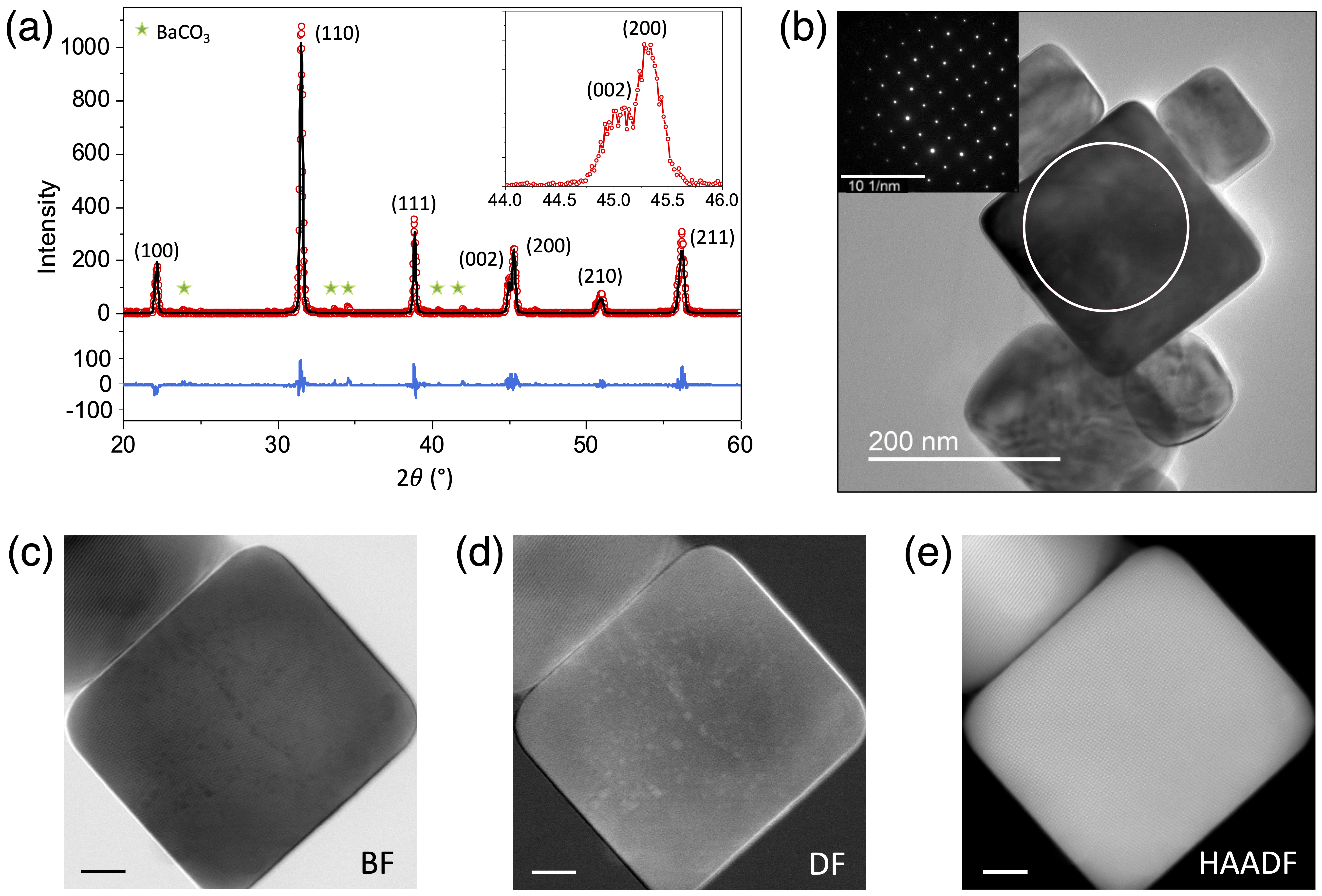}
    \caption{\textbf{Structural characterization of barium titanate nanocubes.} (a) X-ray diffractogram showing all the detected peaks indexed to the tetragonal phase of BTO. The data points observed are represented as red circles, and the calculated intensities are shown as a black solid line. The inset graph shows, near $2\theta=45^\circ$, two separate peaks for (002) and (200) planes evidencing the tetragonal structure. Green stars indicate small peaks associated to the structure of residual BaCO$_3$ crystallites precursor. The solid blue line at the bottom represents the difference between the observed and calculated intensities. (b) TEM image of a small aggregate of nanocrystals. Scale bar: 200~nm. Inset: electron diffraction pattern on the encircled region, consistent with a single-crystal nature. Scale bar: 10~nm$^{-1}$. (c-e) HR-STEM images of another BTO nanocrystal, in bright-field BF (c), dark-field DF (d) and high-angle annular dark-field HAADF (e) modes. Scale bar: 20~nm. BF and DF modes display multiple square-shaped spots within the NC, which may be little pores.}
    \label{fig:XRD_TEM}
\end{figure}

Quasi-cubic shaped barium titanate nanocrystals were synthesized using the solvothermal method described in Bogicevic~\textit{et al.}~\cite{Bogicevic.2015} (Materials and Methods). Supporting Information Figure~S1a shows a large field-of-view scanning electron microscopy image of the as-produced powder, confirming the overall quasi-cubic shape. We first did a structural analysis of this powder, from which we then prepared the dilute suspension for PFM studies of single nanocrystals.

Figure~\ref{fig:XRD_TEM}a displays the powder X-ray diffractogram at the temperature of 293~K showing the characteristic peak splitting of tetragonal phase at twice the diffraction angle $2\theta\approx$45$^\circ$, corresponding to the $(hk\ell)$ Miller indexes $(002)$ and $(200)$, while a purely cubic phase would have a single peak associated to $(002)$. Rietveld refinement of the diffractogram (Materials and Methods) reveals a well-known tetragonal structure (space group symmetry P4mm) for BTO at room temperature (293~K), with $a=b=4.0072\pm\kern -1.8pt0.0005$~\AA \, and $c=4.0339\pm\kern -1.8pt0.0006$~\AA, leading to a lattice parameter ratio $(c/a)_{\rm NC}=1.0067\pm\kern -1.8pt0.0002$, with weighted profile and expected reliability factors $R_{\rm wp}=11.52\%$ and $R_{\rm exp}=6.63\%$ respectively, from which the goodness-of-fit $\chi^2\equiv (R_{\rm wp}/R_{\rm exp})^2=3.02$ is inferred.
Compared to bulk BTO for which $(c/a)_{\rm bulk}=1.0101\pm\kern -1.8pt 0.0002$~\cite{Megaw.1945}, the ratio measured in our NC is significantly smaller by 0.3\% and closer to one, indicating that its lattice is slightly less tetragonal and more cubic than the bulk one, in agreement with other reports on BTO nanocrystals~\cite{Huang.2017ar,Lee.2012}.
We also used transmission electron microscopy (TEM, Materials and Methods) to investigate the crystalline structure at the single particle level. Figure~\ref{fig:XRD_TEM}b shows a TEM image of a small aggregate accompanied with the diffraction pattern of the selected area on the central nanocrystal, which sizes are 211~nm$~\times~$226~nm. The regular pattern obtained is consistent with the monocrystalline nature of the selected NC. 

The reduced tetragonality that we observed in BTO nanocrystals compared with the bulk was reported in several studies and was attributed to different sources. Firstly, one study evidenced by high-resolution scanning TEM (HR-STEM) the presence of a thin shell of cubic symmetry at BTO NC surface~\cite{Zhu.2009}. This observation was further supported by indirect measurements and models~\cite{Lee.2012}. We examined our BTO NC edge with HR-STEM (see Figure~S2) but could not detect a layer of different symmetry. 
Secondly, we estimated to 0.19\% the inhomogeneous strain from Figure~\ref{fig:XRD_TEM}a diffractogram using Williamson-Hall method with a Cauchy peak profile~\cite{Vivekanandan.1989} (see Figure~S1b), which is smaller than the value of 0.26\% recently reported~\cite{Suzana.2023}, but not negligible. This strain is probably due to hydroxyl groups coming from the barium precursor and remaining inside the nanocrystals as point structural defects. The latter would migrate to the surface only at temperatures much higher than those used during solvothermal synthesis.
Other sources of strain are small square-shape pores that we observed in bright-field (Figure~\ref{fig:XRD_TEM}c) and dark-field HR-STEM (Figure~\ref{fig:XRD_TEM}d) of one nanocrystal (with sizes 116~nm$~\times~$128~nm), but not in high-angle annular dark-field which rather shows chemical contrast.
In some particles with rounder morphology we even observed much larger pores (see Figure~S2c,d). We reported such pores in a previous study on the same sample~\cite{Bogicevic.2015}, and showed evidences that they result from the merging and restructuration, upon temperature increase above 200$^\circ$C, of nanotori, which are the primary structure formed at low temperature.

As these defects may impact the ferroelectric properties, we further conducted Raman spectroscopy on BTO powder, at temperatures varying between 80~K and 540~K. In the centro-symmetrical paraelectric cubic phase, BTO is Raman inactive, contrary to the tetragonal phase, which symmetry leads to height Raman optical-phonon active modes\cite{DiDomenico.1968}. We monitored the area variation of the narrowest active Raman peak (305~cm$^{-1}$ wavenumber, at 300~K) with temperature (see Figure~S3a), as in Begg \textit{et al.}~\cite{Begg.1996} and observed a marked decrease at $395\pm 2$~K (Figure~S3b) in agreement with the commonly accepted BTO Curie temperature of 120$^\circ$C. This observation indicates that, despite a reduced tetragonality compared with the bulk, the BTO NC we synthesized keeps the hallmark of the bulk BTO ferroelectric phase transition.
Our objective was then to examine the ferroelectric domain texture of single BTO nanocrystals, and to this aim we performed vector piezoresponse force microscopy~\cite{Kalinin.2006}.

\subsection{Piezoresponse displacement measurements}
PFM allows nondestructive imaging and control of ferroelectric domains at the nanoscale via bias-induced converse piezoelectric effect mechanical deformations. It has gained interest in helping the development of  nanoferroics~\cite{Zhang.2021des,Gruverman.2019}. The basic principles of PFM are recalled in Supporting Information Text~S1 and for example in ref~\cite{Kalinin.2007}. 
PFM is a scanning probe microscopy technique operating in contact mode, which makes the study of individual nanocrystals challenging, as they must be firmly immobilized on a conductive substrate, while having their upper surface directly exposed to the conductive tip.

\begin{figure}[h!]
    \centering
    \includegraphics[width=\textwidth]{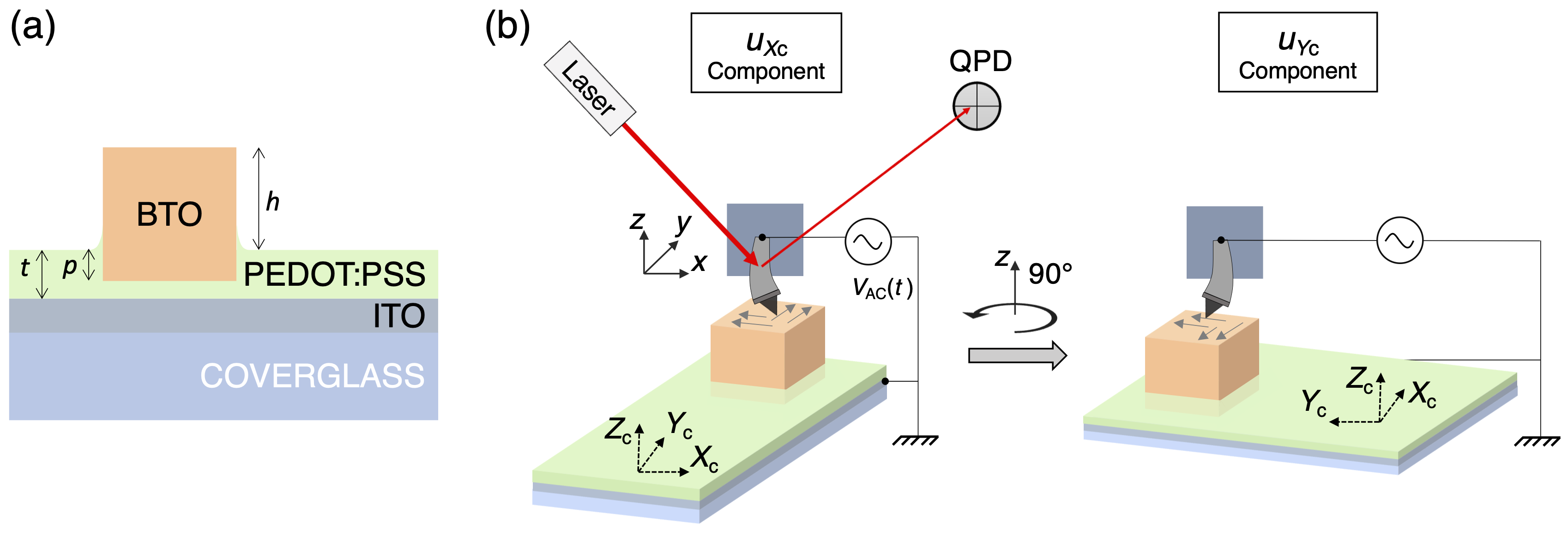}
    \caption{\textbf{Experimental configuration of lateral piezoforce response measurement on a single BTO nanocrystal}. (a) Immobilization of BTO NC onto an ITO-coated coverglass with a thin film of PEDOT:PSS conductive polymer. The dimensions shown are: $t$, the polymer thickness, $h$, the height of the particle emerging part above the polymer layer of the particle, and $p$, the penetration depth. With $d$, the particle size, we have $p=d-h$ and $p<t$. (b) Schematics of lateral PFM measurement. $(x,y,z)$ and $(X_{\rm c},Y_{\rm c},Z_{\rm c})$ are the laboratory and crystal-attached frames, respectively. The measurement consists in applying an AC-voltage $V_{\rm AC}(t)=V_{\rm AC}\cos(2\pi f_{\rm d}t)$ to the tip, with $f_{\rm d}$ being the driving frequency, set close to the tip contact resonance, and $V_{\rm AC}=1.5$~V. The PFM tip, aligned along the $y$ axis, is scanned along the $x$ axis, which is colinear to $X_{\rm c}$ axis initially (left configuration). During scanning, the nanocrystal in-plane polarization domains (displayed by gray arrows) induce shear displacements of the tip along the $x$ axis, oscillating in phase or out of phase with $V_{\rm AC}(t)$, depending on the orientation of the projection of the polarization along the $X_{\rm c}$ axis. The resulting displacement amplitude $u_{X_{\rm c}}$ is inferred from the deflection of the PFM laser as measured by the quadrant photodiode (QPD). To reduce electrostatic artifacts, we set the laser beam to hit the cantilever at the electrostatic blind spot (ESBS, see main text). To access the displacement $u_{Y_{\rm c}}$ along the $Y_{\rm c}$ axis, we rotate the sample by 90$^\circ$ counterclockwise around the laboratory $z$ axis, and repeat the scan along the $x$ axis (right configuration; the laser and QPD are not shown for the sake of clarity).}
    \label{fig:Experimental_config}
\end{figure}

For such a study we prepared a sample with isolated NC attached to a conductive substrate made of indium tin oxide (ITO)-coated coverglass, with a marking grid on top of the coating, to facilitate the correlation between scanning electron microscopy prelocalization of isolated NC and their subsequent study by PFM. The ITO-coated coverglass with the grid is first covered with a thin layer of a conductive polymer (Poly(3,4-ethylenedioxythiophene)-poly(styrenesulfonate), PEDOT:PSS), on top of which a dilute aqueous suspension of the NC is spincoated (Materials and Methods), as displayed on Figure~\ref{fig:Experimental_config}a. We optimized the various parameters to achieve a polymer thickness in the range of 40-50~nm (see Figure~S4a). 
Thanks to a good wettability of BaTiO$_3$ by water~\cite{Li.2014}, the concave meniscus (visible on Figure~S4d) forming on the edges of the nanoparticles drags them down by surface tension, and after drying the polymer layer in an oven, about two-thirds of the particle height emerges and can be directly contacted by the PFM conductive tip.

Figure~S5a shows a typical low magnification SEM image of the sample with an apparent cell of the numbered grid. To conduct the PFM study, we first selected and located isolated NC, like the one at the center of Figure~S5b, and characterized their size distribution, that fell in the range of 120-210~nm, as shown on Figure~S6a. Moreover,  Figure~S6b indicates that about 65\% of these NC have an planar aspect ratio larger than 0.95 in the sample plane, which is a good indication of close to perfect cubic shape.

PFM measurements were conducted according to the experimental configuration presented in Figure~\ref{fig:Experimental_config}b.
We first took care of minimizing artifacts that could stem from electrostatic charge accumulation~\cite{Killgore.2022}. We evidenced such parasitic effects when using conventional deflection laser spot setting at the cantilever tip. These effects include the observation on the conductive and nonferroelectric surface of the ITO of an apparent contact resonance of the cantilever (see Figure~S7a), an hysteresis loop (Figure~S7b), and charge accumulation in freshly ``written'' regions (Figure~S8a,b). We were able to strongly reduce these non-electromechanical artifacts by implementing the recently published electrostatic blind spot (ESBS) settings~\cite{Killgore.2022} as detailed in the Supporting Information Data S2.2 and shown on Figure~S7c,d and Figure~S8c-f.

Once these parasitic effects were minimized, we could more reliably map the lateral PFM two-dimensional displacement field (LPFM) of the top surface of immobilized NC. 
We did not apply a DC bias voltage while acquiring PFM data.
As the PFM cantilever scans only along the direction $x$ of the laboratory frame, the sample must be rotated by 90$^\circ$ around the vertical $z$ laboratory axis, as shown on Figure~\ref{fig:Experimental_config}b, to acquire the $Y_{\rm c}$ LPFM component after the $X_{\rm c}$ one, $(X_{\rm c},Y_{\rm c},Z_{\rm c})$ being a frame of reference attached to the NC of interest.
\begin{figure}[h!]
    \centering
    \includegraphics[width=\textwidth]{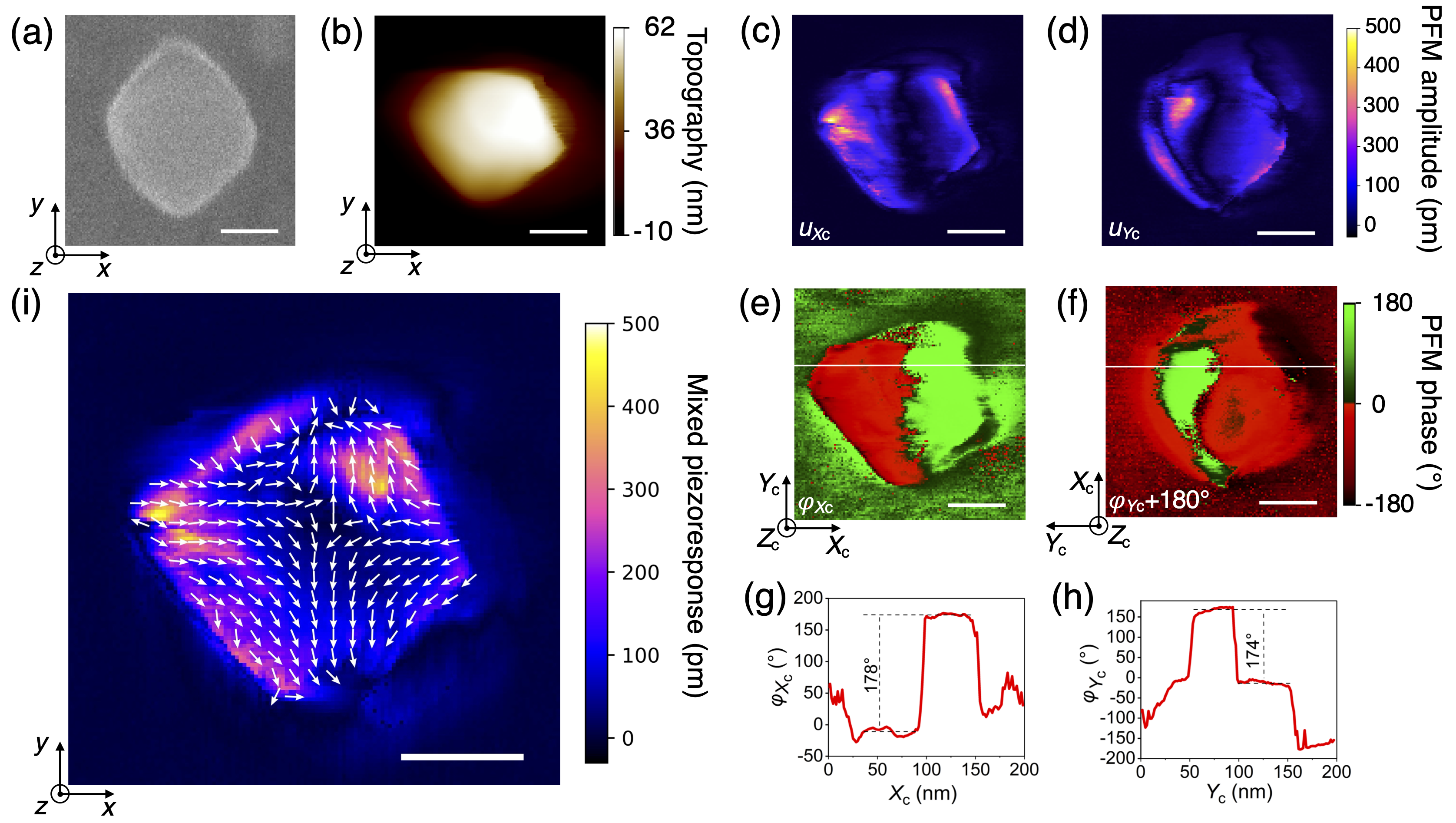}
    \caption{\textbf{Lateral piezoresponse displacement mapping on a single BTO NC}. (a) SEM scan of the investigated particle, with an approximately square shape and size of 125~nm. The laboratory frame is denoted $(x,y,z)$, with (O$z$) perpendicular to the substrate. We scan the cantilever along the $x$ axis. (b) AFM height measurement (contact mode) of the same BTO NC oriented as shown in (a). (c,d) Amplitudes $u_{X_{\rm c}}$ and $u_{Y_{\rm c}}$ of lateral piezoresponses along the $X_{\rm c}$ and $Y_{\rm c}$ axis respectively of the frame $(X_{\rm c},Y_{\rm c},Z_{\rm c})$ attached to the nanocrystal and confounded with $(x,y,z)$ before sample rotation. To record $u_{Y_{\rm c}}$ the sample was rotated 90$^\circ$ counterclockwise around the $z$ axis. The amplitude of the electric field modulation applied was 1.5~V. (e,f) Lateral piezoresponse phase associated to $X_{\rm c}$- and $Y_{\rm c}$ axis respectively. (g,h) Cross sections of the (e) and (f) phase maps. (i) Mixed lateral piezoresponse field $\boldsymbol{u}^{\rm m}(u_{X_{\rm c}}\cos(\varphi_{X_{\rm c}}-\varphi_{\rm offset}), u_{Y_{\rm c}}\cos(\varphi_{Y_{\rm c}}-\varphi_{\rm offset}))$, displayed by a superimposed norm (color code) and direction maps (array of fixed-length arrows at the nodes of a 7.8~nm pitch grid). Note that the lateral displacement values are not absolute, as we did not perform any calibration. Scale bars (all scans): 50~nm.}
    \label{fig:LPFM}
\end{figure}

We implemented this procedure to map the LPFM of four nanocrystals, setting the voltage modulation amplitude applied to the tip at $V_{\rm AC}=1.5$~V for all of them. Figure~S9 shows a typical cantilever contact resonance curve, used to set the driving frequency close to its maximum.
Figure~\ref{fig:LPFM} exemplifies LPFM measurements in one of the four NC, whose lateral dimensions measured using SEM (Figure~\ref{fig:LPFM}a) are $125\text{~nm}\times125\text{~nm}$ ($d=125$~ nm). The height of the portion of the NC emerging from the PEDOT:PSS layer, as measured by AFM, is $h\approx 72$~nm (Figure~\ref{fig:LPFM}b). Assuming a cubic shape, the depth of penetration of the particle in the polymer layer is then $p\equiv d-h\approx 53$~nm (using Figure~\ref{fig:Experimental_config}a notations), which is consistent with Figure~S4a thickness estimate. 

Figure~\ref{fig:LPFM}c-h then display LPFM measurements carried out on this NC, for which the electric field applied by the tip (along the laboratory $z$ axis) has an amplitude $E^{\rm exp}_z=120$~kV/cm across the supposedly 125~nm-sized particle.
Figure~\ref{fig:LPFM}c shows a map of $u_{X_{\rm c}}$, which is the amplitude of the voltage-induced local lateral deformation along the $X_{\rm c}$ direction.
Figure~\ref{fig:LPFM}e shows the map of the phase difference $\varphi_{X_{\rm c}}$ between $V_{\rm AC}$ applied to the tip and the NC piezoresponse along the $X_{\rm c}$ direction.
In order to confirm that these displacement and phase values are not artefactual, we applied our measurement protocol to a reference sample consisting of a $Y_{\rm c}$-cut periodically-poled lithium niobate rod (kindly prepared for us by Prof. Mathieu Chauvet, Femto-ST, Besançon, France) that has only in-plane ferroelectric domains in the $Y_{\rm c}$ facet and only out-of-plane in the $Z_{\rm c}$ (poling direction) facet. Figure~S10 displays LPFM measurements carried out on the $Y_{\rm c}$ facet of this reference sample showing unambiguously domains with opposite polarities, as evidenced by the phase shift close to 180$^\circ$ across the domain walls. Moreover, the displacement amplitude of $\approx 120$~pm is smaller than the one we measured for the nanoBTO (about 500~pm for the particle reported in Figure~\ref{fig:LPFM}), in agreement with the smaller $d_{15}$ lateral piezoelectric coefficient of lithium niobate~\cite{Weis.1985} ($d_{15}^{\rm LN}\approx 70$~pC/N) compared to the one of BTO~\cite{Berlincourt.1958} ($d_{15}^{\rm BTO}\approx 270$~pC/N). The ratio of the displacement amplitude (linearly corrected for difference in driving voltage amplitude, 2~V for PPLN and 1.5~V for BTO) is $\approx 5.6$, which is in good quantitative agreement with that of $\approx 3.9$ for $d_{15}$ ratio, considering that we did not use the same tip for both measurements.

As our LPFM measurements led to consistent results on the reference sample, we considered that the data we recorded on BTO NC with the same methodology correspond to a real piezoforce response.

Then, to access LPFM measurement along the $Y_{\rm c}$ direction of the nanocrystal, the sample was rotated counterclockwise by $90\pm\kern -1.8pt 2$$^\circ$, around the laboratory $z$ axis. Figure~\ref{fig:LPFM}d and Figure~\ref{fig:LPFM}f show the amplitude $u_{Y_{\rm c}}$ and phase $(\varphi_{Y_{\rm c}}+180\text{$^\circ$})$ maps respectively, along the $Y_{\rm c}$-direction.
Note that the phase actually measured in the rotated-sample orientation is $\varphi_{Y_{\rm c}}+180$$^\circ$ to account for the fact that the $Y_{\rm c}$ axis, after rotation, points in the opposite direction of the original $X_{\rm c}$ axis.
We observed a small instrumental phase offset, $\varphi_{\rm offset}=-10\pm\kern -1.8pt 2$$^\circ$, for the in-phase piezoresponse (refer to Materials and Methods for its determination), which we arbitrarily chose to be the phase of the in-phase domain (with a positive $X_{\rm c}$ component).
The cross-sections of Figure~\ref{fig:LPFM}e,f evidence a phase difference of $\approx180^\circ$ between the red and green regions, revealing the presence of in-plane electric polarization domains with opposite orientations along the scanning direction.

We finally inferred the mixed lateral piezoresponse displacement field $\boldsymbol{u}^{\rm m}$ components, defined as $u^{\rm m}_{X_{\rm c}}\equiv u_{X_{\rm c}}\cos(\varphi_{X_{\rm c}}-\varphi_{\rm offset})$ and $u^{\rm m}_{Y_{\rm c}}\equiv u_{Y_{\rm c}}\cos(\varphi_{Y_{\rm c}}-\varphi_{\rm offset})$, displayed in Figure~\ref{fig:LPFM}i.
We observe that the displacement directions close to the nanocrystal edges are either parallel or orthogonal to these edges, and that these orientations propagate from the surface to the center.

The same behavior was observed for the lateral PFM applied to the three other individual nanocrystals investigated (see Figures~S12 to S14), highlighting the fact that our observations are reproducible.
The measured lateral responses display non-homogeneous converging deformation fields that reflect complex underlying ferroelectric textures, and that we further investigated by simulations, as discussed later. Note that similar converging deformation fields were reported in single bismuth iron oxide dots of cylindrical shape, nanostructured as an array in a thin layer~\cite{Li2017}.
\begin{figure}[h!]
    \centering
    \includegraphics[width=\textwidth]{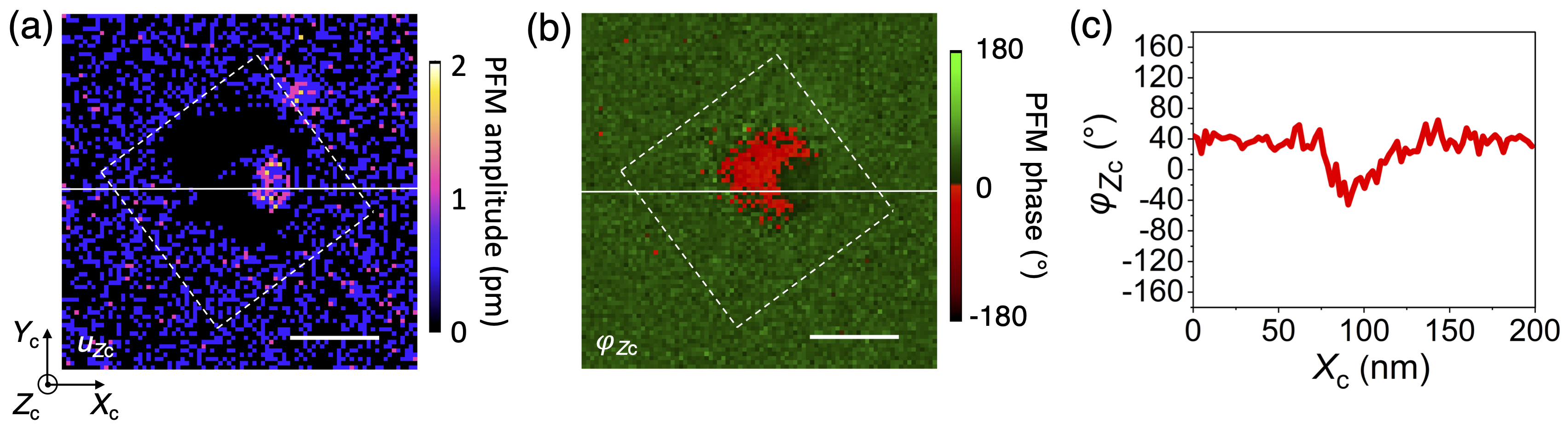}
    \caption{\textbf{Vertical piezoresponse of the same nanocrystals as the one of Figure~\ref{fig:LPFM}, which footprint is marked by the white dashed line square}. (a,b) PFM amplitude (a) and phase (b) maps showing no significant contrast. (c) Cross section of the phase image along the solid white line. Scale bars (all images): 50~nm.}
    \label{fig:VPFM}
\end{figure}

We also measured the vertical piezoresponse (VPFM) that is shown in Figure~\ref{fig:VPFM}, and detected no significant vertical piezoresponse amplitude compared to the lateral one as $u_{Z_{\rm c}}^{\rm max}/u_{X_{\rm c}}^{\rm max}\approx 4\times 10^{-3}$ (Figure~\ref{fig:VPFM}a).
We observe a phase difference of $\approx -80$$^\circ$ relative to the surrounding, in the exact same region where there is no displacement amplitude at all. We therefore suspect that this phase value has no physical meaning.
In order to validate our protocol for VPFM measurements, the out of plane polarization of the $Z_{\rm c}$ oriented previously mentioned PPLN reference rod was studied  (Figure~S11).

In order to interpret these PFM data, we carried out ferroelectric phase field simulations, first without any applied external electric field, to get the equilibrium ferroelectric domain distribution, and then in condition of PFM measurement, with an external applied electric field which amplitude is of the order of magnitude of the one used in the experiment.

\subsection{Phase field simulations}
We considered a cubic-shape BTO nanocrystal of 100~nm size, embedded in air and used the FERRET module~\cite{Mangeri2017} of the Multiphysics Object Oriented Simulation Environment (MOOSE), an open-source software maintained by Idaho National Laboratory~\cite{lindsay2022moose}. FERRET is a finite element implementation of a ferroelectric phase field model, which we utilise here to capture the polar and elastic equilibrium state of BTO nanocrystals, as well as their response to static electric fields. 
In the experiment, one facet of the particle is in contact with a conductive medium. However, as shown in Prosandeev and Bellaiche~\cite{Prosandeev2007}, asymmetric screening with a metal electrode does not destroy the domain structure in ferroelectric thin films, and the simulations conducted probably extend to the experimental case of the bottom facet surface attached to a conductive layer with the rest of the nanocube exposed to air.
Moreover, due to computational limitations, we did not explore NCs larger than 100 nm. However, our set of data for smaller sizes indicates that the dielectric constant, a probe of the domain structure and its response to applied electric fields, converges for particle sizes larger than 50 nm (see Figure S15). This implies that the domain structure is likely to be very similar for NCs size beyond 50 nm, and thus, we expect that particles slightly larger than 100 nm like the one we investigated experimentally, have similar physical responses.

\paragraph{Equilibrium domain structure.}
\begin{figure}[h!]
    \centering
    \includegraphics[width=\textwidth]{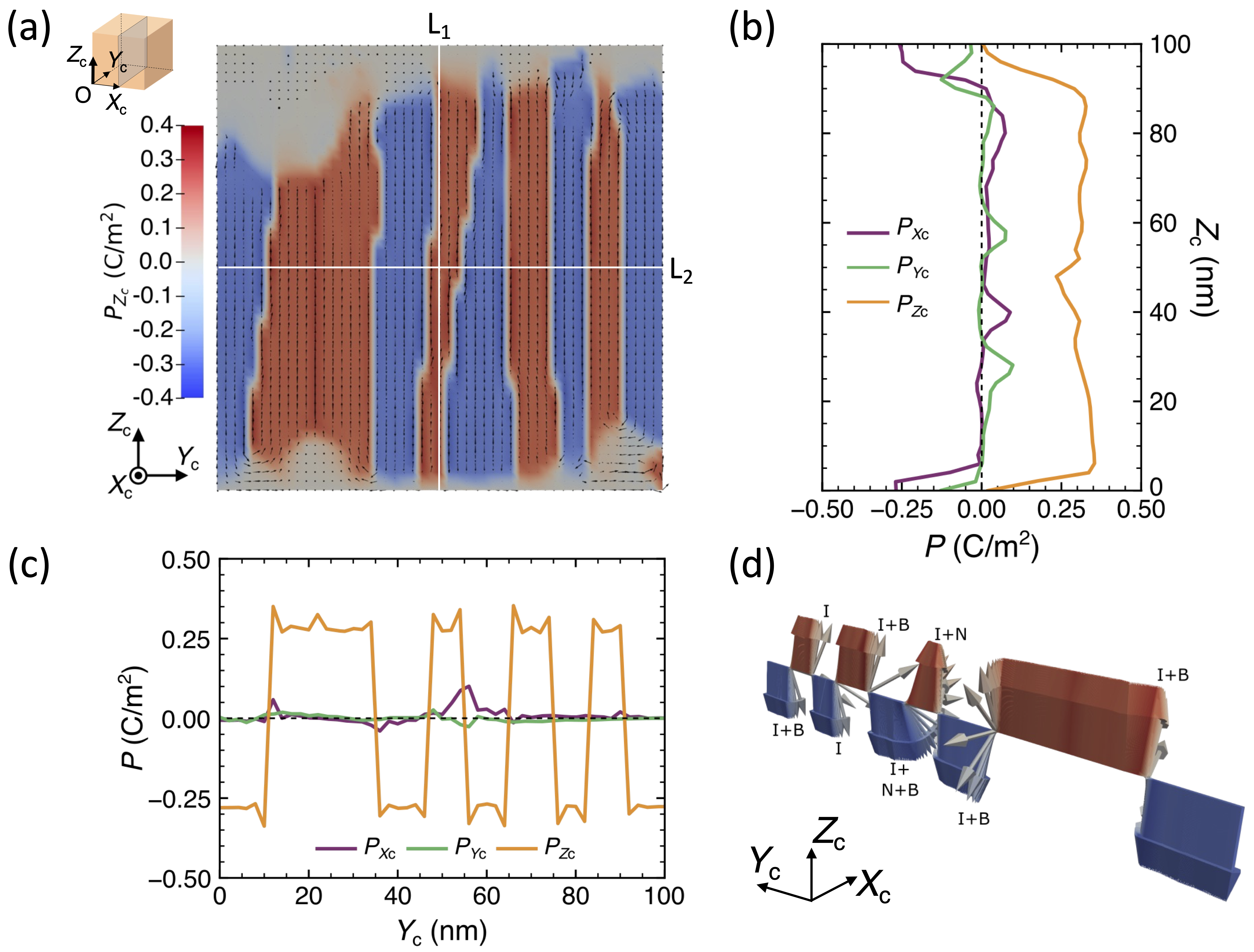}
    \caption{\textbf{Polarization components distribution at equilibrium in the middle plane $X_{\rm c}=50$~nm of a 100~nm sized cube of BTO}. (a) Planar distribution of the $P_{Z_{\rm c}}$ component defining the main (O$Z_{\rm c}$) polar axis ([001] direction). The alternation of red and blue shadings associated with positive and negative $P_{Z_{\rm c}}$ components, respectively, illustrates the multiple 180$^\circ$ domains. The black arrow glyphs map the polarization field 3D orientation every 2~nm. Note the small thickness of the surface domains (top and bottom surfaces) with $P^{\rm eq}_{Z_{\rm c}}\approx 0$. (b) Profiles of the polarization components along the line L$_1$ ($X_{\rm c}=50$~nm, $Y_{\rm c}=50$~nm, $Z_{\rm c}$) showing the presence of surface domains with polarization perpendicular to the main (O$Z_{\rm c}$) polar axis. (c) Profiles of the polarization components along the line L$_2$ ($X_{\rm c}=50$~nm, $Y_{\rm c}$, $Z_{\rm c}=50$~nm) showing the partial Bloch and Néel nature of some domain walls. (d) 3-dimensional representation of the polarization vector along the same line as in (c), with labeled types of domain walls (B stands for Bloch, N for Néel and I for Ising; I+B, I+N and I+B+N indicate a mixed Ising-Bloch, mixed Ising-Néel and mixed Ising-Bloch-Néel character respectively).}
    \label{fig:Polar_equilibrium_distribution}
\end{figure}

We first determined the equilibrium polarization and displacement fields, $\boldsymbol{P}^{\rm eq}(\boldsymbol{r})$ and $\boldsymbol{u}^{\rm eq}(\boldsymbol{r})$ by minimising the energy of a BTO nanocube with random and small initial polarization and mechanical displacement field components.
Figure~\ref{fig:Polar_equilibrium_distribution}a shows that minimization of the free energy of the system leads to a multidomain structure consisting of domains polarized upward and downward along the (O$Z_{\rm c}$) axis, which we now refer to as the polar axis [001]. 
We estimate that the core of the BTO NC made of upward and downward tetragonal domains along (O$Z_{\rm c}$) represents 60\% of the fraction of the volume of the NC.
From the polarization profiles along lines L$_1$ and L$_2$ (Figure~\ref{fig:Polar_equilibrium_distribution}b,c), we inferred a polarization component amplitude $P_{Z_{\rm c}}=\pm 0.265$~C/m$^2$ in these up and down domains, which is consistent with what has been measured~\cite{Wieder1955} or calculated~\cite{Hlinka2006PhenomenologicalMO,Wang2010} in the tetragonal ferroelectric phase of barium titanate.

Looking at the domain walls separating up and down domains in Figure~\ref{fig:Polar_equilibrium_distribution}c polarization profile, we observe that they are mostly of the Ising-type, \textit{i.e.} the transition is accomplished by modulating the $P_{Z_{\rm c}}$ component. We do see, however, that some walls retain partial Bloch characteristics (rotation of the polarization in planes parallel to the domain wall, \textit{i.e.} showing non-vanishing $P_{X_{\rm c}}$ component, for example at $Y_{\rm c}=$11, 35, 55 and 90~nm) or Néel characteristics (rotation of the polarization in a plane orthogonal to the domain wall, \textit{i.e.} with non-zero $P_{Y_{\rm c}}$, for example at $Y_{\rm c}=$45 and 55~nm), as schematized on the polarization profile line L$_2$ in Figure~\ref{fig:Polar_equilibrium_distribution}d.

Note that, in surface planes along the polar axis like $X_{\rm c}=0$~nm, the polarization aligns up and down along (O$Z_{\rm c}$) like in the bulk (see Figure~S16a) and the domain walls are also of Ising type with partial Néel characteristics (Figure~S16b).

\begin{figure}[h!]
    \centering
    \includegraphics[width=\textwidth]{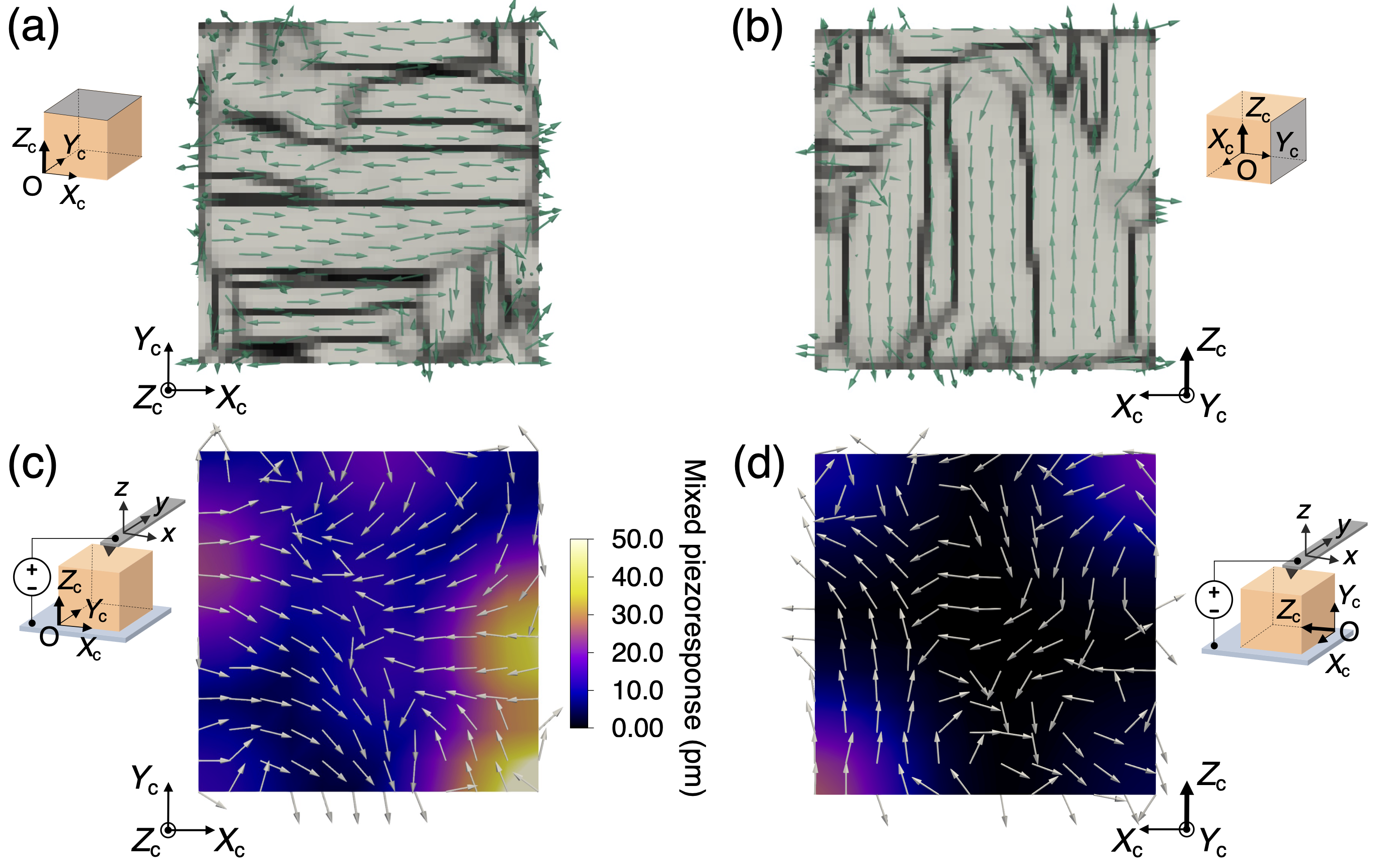}
    \caption{\textbf{Simulated in-plane polarization texture and piezoresponce of 100~nm sized BTO cube, with polar axis along (O$Z_{\rm c}$)}. (a) Polarization gradient in the $Z_{\rm c}=100$~nm (001) plane at equilibrium (at the temperature of 293~K), with light gray indicating domains and dark grey/black indicating domain walls. The solid arrow glyphs (green) represent the orientation of the polarization vector. (b) Polarization gradient in the $Y_{\rm c}=100$~nm  (010) plane.
    (c) In-plane mixed piezoresponse in the $Z_{\rm c}=100$~nm plane, for a crystal oriented with the $Z_{\rm c}$ axis parallel to the laboratory $z$ axis, in response to a field of amplitude $E_z=150$~kV/cm. The solid arrow glyphs indicate the piezoresponse vector orientation at the surface. The color map corresponds to a gaussian resampling of the simulation values of the piezoresponse displacement to better match experiments (kernel standard deviation of $\approx 20$~nm, corresponding to the experimental imaging resolution). The largest calculated displacement, located at the bottom-right corner ($Y_{\rm c}=0$~nm and $X_{\rm c}=100$~nm) is 75~pm, and the displacement in the center of the cube is 4~pm. (d) In-plane mixed piezoresponse in the $Y_{\rm c}=100$~nm plane, for a crystal oriented with $Z_{\rm c}$ axis parallel to laboratory $x$ axis, in response to a field applied along $z$ axis (amplitude $E_z=150$~kV/cm, very close to the field amplitude applied in the PFM measurements).}
    \label{fig:simul_domains_and_PFM_displacement}
\end{figure}

Let us now focus our attention on the facets of the NC perpendicular to the polar axis, $Z_{c}=0$ and $Z_{c}=100$~nm. Interestingly, the $P_{Z_{\rm c}}$ component of the polarization field vanishes near the surface of these facets (see Figure~\ref{fig:Polar_equilibrium_distribution}a,b). Instead, the polarization lies in the plane of these facets, as shown in Figure~\ref{fig:simul_domains_and_PFM_displacement}a and Figure~S16c (facet $Z_{c}=100$~nm), with mostly tetragonal domains polarized along the (O$X_{\rm c}$) direction (representing about 16\% of the total NC volume), as well as a lower number of domains polarized along the (O$Y_{\rm c}$) direction (about 7\% of the NC volume).
These surface domains extend at an average depth of 8~nm (see Figure~\ref{fig:Polar_equilibrium_distribution}a,b) from the surface of the $Z_{\rm c}=0$ and $Z_{\rm c}=100$~nm facets, and they also display Ising domain walls with Bloch and/or Néel characteristics (Figure~S16d) like in the core of the nanocrystal (Figure~\ref{fig:Polar_equilibrium_distribution}d).

Rotation of the polarization from (O$Z_{\rm c}$) direction to either (O$X_{\rm c}$) or (O$Y_{\rm c}$) one allows the system to prevent a strong electrostatic depolarizing field. Indeed, the averaged planar electrostatic energy density in the equilibrium structure is mostly zero (see Figure~S17). Depolarizing fields, which appear when polarization-induced bound charges are imperfectly screened at an interface, are well known in ferroelectric thin films~\cite{Mehta1973,Kopal1997}. They are responsible for the emergence of flux closure patterns~\cite{Aguado-Puente2008,Prosandeev2007} in which the polarization field rotates from the out-of-plane direction to lie in the plane of the thin film surface, similarly to our NCs, as well as for the appearance of more exotic polar textures such as polar vortices~\cite{Hong2017}, which we do not observe here.
Rotation of polarization on surfaces normal to (O$Z_{\rm c}$) thus allows the NC to sustain uniformly polarized domains, at the cost of an increase in electromechanical energy near the surfaces. Nonetheless, this is counteracted by the release of elastic energy by the $90^{\circ}$ domain walls near the surface (see Figure~S17).

\paragraph{Lateral piezoresponse force response simulation.}

Starting from the equilibrium solution described in the previous paragraph, we now apply a finite electric field to simulate PFM experimental conditions. The BTO nanocrystal is then relaxed under this applied electric field, leading to new equilibrium values of the polarization and mechanical displacement fields. 

Let us first consider the case when an electric field of amplitude $E^{\rm sim}_z=150$~kV/cm is applied along the polar axis (O$Z_{\rm c}$) direction of the BTO nanocrystal. The difference between the zero-field and finite-field mechanical displacement fields gives us an estimate of the electromechanical response of the nanocube. We show, in Figure~\ref{fig:simul_domains_and_PFM_displacement}c, the resulting electric-field-induced mixed piezoresponse in the ``top'' facet perpendicular to the polar axis ($Z_{\rm c}=100$~nm facet). 
We observe qualitatively similar LPFM data of Figure~\ref{fig:LPFM}i, Figure~S12f, S13f and S14f, with local regions of pronounced in-plane displacement at the corners and edges of the facet. We also observe a low to vanishing in-plane piezoelectric displacement in the inner region of the facet, as evidenced experimentally as well.

To investigate the other experimental configuration of the nanocube having its polar axis (O$Z_{\rm c}$) lying in the substrate plane, parallel to the laboratory plane $(xy)$, we also simulated the application of an electric field along the (O$Y_{\rm c}$) axis using the same equilibrium condition as depicted in Figure~\ref{fig:Polar_equilibrium_distribution}. 
The resulting mixed piezoresponse simulation, presented on Figure~\ref{fig:simul_domains_and_PFM_displacement}d, exhibits lower (by one order of magnitude in the center of the facet and by a factor of 2 to 5 at the edges) and contrast than the simulation for the field applied along (O$Z_{\rm c}$). 
It is well known that the presence of 90$^\circ$ twin domain walls contribute to the piezoelectric response of BTO~\cite{Wada2006,Hlinka2009,Ghosh2014} and can strongly enhance it. Indeed, contrary to the (O$Y_{\rm c}$) facet, the (O$Z_{\rm c}$) possesses multiple 90$^{\circ}$ domain walls, at a distance of about 8~nm from the surface, due to a 90$^{\circ}$ rotation of the polarization from a $Z_{\rm c}$ axis aligned orientation (as displayed for $X_{\rm c}=0$~nm facet in Figure~S16a) to an in-$(X_{\rm c},Y_{\rm c})$ plane lying one. This may explain the larger simulated in-plane electromechanical response when the electric field is applied along the $Z_{\rm c}$ axis (Figure~\ref{fig:simul_domains_and_PFM_displacement}c).

We note that the magnitude of the maximum in-plane displacement observed experimentally (Figure~\ref{fig:LPFM}i) is one order of magnitude larger than that obtained from the simulation (Figure~\ref{fig:simul_domains_and_PFM_displacement}b). Apart from the fact that we did not calibrate PFM lateral displacement, two other factors may contribute to this difference. First, in the experiment the amplitude of the electric field effectively applied to the nanocrystal is not known precisely because the PFM geometry results in a pronounced inhomogeneous radial distribution with a larger field value compared to the one of the uniform field applied in the simulations (considering the same applied voltage). Second, in the simulations the applied electric field is static, while the PFM tip applies an AC voltage oscillating at a frequency set close to the electromechanical resonance, hence enhancing the displacement.

Finally, we carried out the phase field simulations considering perfectly cubic geometry, while the synthesized nanocrystals exhibit round edges (with radius of curvature of about 15~nm for Figure~\ref{fig:XRD_TEM}c-e NC which size is $\approx 120$~nm). We could wonder if such non-perfectly cubic shape could impact the simulations. However, the influence of NC shape on the phase field predictions was investigated by Pitike \textit{et al.}\cite{Pitike.2018} who reported very similar qualitative polarization textures for perfectly cubic or nearly cubic NC. This behavior consists in particular in a transition from a single polarization domain to multiple domains, that takes place at sizes of about 4~nm for cubic and 2~nm for nearly cubic, both far smaller than the NC sizes we studied. The low sensitivity of the polarization texture to the deviation from perfectly cubic shape for NC size $\approx 100$~nm is also supported by the fact that we observed the same qualitative polarization texture for the four particles investigated experimentally (see Figure~\ref{fig:LPFM}i, and supporting Figures~S12f to S14f).

\section{Conclusion}
It is now well established theoretically and experimentally, that ferroelectric materials at the nanoscale exhibit exotic polarization texture, leading to unconventional properties. The case of 0-dimension ferroelectric nanocrystals, in particular isolated ones, has been far less studied experimentally, despite the opportunity it offers to bring not yet reported properties due to their large surface to volume ratio.
In this work we considered isolated barium titanate nanocrystals immobilized on a conductive surface and exposed to air. Using a solvothermal approach~\cite{Bogicevic.2015} we were able to synthesize NC of a well-defined cubic shape and 160~nm average size, that conserve the tetragonal structure of BTO at room temperature (293~K), are individual monocrystallites (as observed by TEM) and exhibit the expected transition to a cubic symmetry phase above 120$^\circ$C, the known Curie temperature of the ferro- to para-electric phases. 
We then investigated, at room temperature, the inverse piezoelectric properties of such pristine individual NC by piezoresponse force microscopy. To this end, we developed a methodology that combines the firm attachment of the NC to the substrate with a conductive polymer and the fine adjustment of the deflection laser in a position along the PFM cantilever that strongly reduces electrostatic artifacts~\cite{Killgore.2022}. This strategy allowed repeated measurements on the same particle without risk of being dragged with the PFM tip and provided a high confidence in the crystal deformation fields recorded. 
We observed that none of the NC showed any vertical PFM displacement, while all showed lateral displacements. We reconstructed mixed piezoresponse fields from the 2D LPFM recording of the same NC for two orthogonal orientations relative to the cantilever (attached to the laboratory frame). 
All the NC studied showed an inhomogeneous LPFM field of larger amplitude along the edges, a minimum at the center with deformations that tend to be oriented towards this lower-amplitude central region.

To check whether our observations were consistent with the underlying texture of the volume polarization, we conducted phase field simulations~\cite{Mangeri2017,Co2021}.
These simulations revealed that in the equilibrium state at the temperature of 293~K, a polar axis (O$Z_{\rm c}$) with up and down 180$^\circ$ domains emerges in the core, accompanied with 90$^\circ$ domains on the two surfaces normal to this axis in a layer of an average thickness of 8~nm. In this layer, there is a dominant polarization orientation in the plane with mostly 180$^\circ$ and a few 90$^\circ$ domains (on the edges). Hence, the phase field simulations predict that none of the cube facets possess an out-of-plane polarization component, which is consistent with the PFM observations. We investigated the nature of the domain walls, as predicted by the simulation, and found that they are mostly of the Ising type with Bloch and/or Néel characteristics.
We then simulated the LPFM field by calculating the difference between the equilibrium position of atomic displacement in the zero electric field and in the presence of an amplitude-similar electric field as applied during the PFM measurement. The resulting simulated LPFM 2D-maps strongly resemble the experimental ones, with maximum displacement amplitudes along the edges and in the corners, and displacement vectors mostly following the gradient of amplitude.
To further consolidate the quantitative reliability of our measurements, we envision to apply more advanced PFM protocols, in particular dual-frequency resonance-tracking~\cite{Rodriguez.2007}.

The results of our study challenge the interpretation of second harmonic generation (SHG) by $\approx 100$~nm-sized BTO nanocrystals, implicitly considered composed of a single static polarization domain in all reported studies~\cite{Karvounis.2020}, eventually bearing a cubic symmetry shell\cite{Rendón-Barraza.2019}. As the 180$^\circ$ domains dominate the core and surface of the NC in a balanced up- and down-fraction, we expect that it yields a destructively interfering SHG field, leaving open the origin of the observed nonlinear contrast. 
Correlative SHG and PFM investigations may contribute to resolve this issue.
Finally, ferroelectric matrices doped with luminescent ionic species reporters have also been considered for sensing applications, including temperature~\cite{Mahata.2020} and electric fields~\cite{Hao.2011,Sun.2017}. The complex polarization structure of BTO nanocrystals that we revealed in our study should be considered to design optimal ferroelectric-based nanosensors.

\section{Materials and Methods}

\subsection{Sample preparation}
\paragraph{BTO nanocube synthesis and characterization.} Barium titanate nanocrystals were synthesized using the solvothermal method described in Bogicevic~\textit{et al.}~\cite{Bogicevic.2015}. Briefly, in a solvent made of volumic proportions of 60\% ethanol and 40\% deionized water, we mixed, in a nitrogen environment, solid Na$_2$Ti$_3$O$_7$ titanium and soluble Ba(OH)$_2\cdot 8$H$_2$O barium precursors. We then heated the mixture at 250$^\circ$C in an autoclave under autogeneous pressure for 24~hours. We selected a Ba/Ti atomic ratio of 1.6 as it leads to BTO nanocrystals of mostly cubic shape and relatively homogeneous size, as can be observed on Figure~S1a SEM image, in the range of 100-350~nm.

XRD measurement (Figure~\ref{fig:XRD_TEM}b) was performed at 293~K using the Cu~K$\alpha$1 ($\lambda$= 1.5418 \AA) and Cu~K$\alpha$2 radiations ($\lambda$= 1.5444 \AA) of a diffractometer (D2 Phaser from Bruker AXS GmbH, Germany) and collecting data in steps of 0.02$^\circ$. We then extracted the structural model with DIFFRAC.EVA (Bruker) software, and performed Rietveld refinement of the structural model using the WinPlotr/FullProf package~\cite{Rodríguez-Carvajal.1993} with peak shape described by a pseudo-Voigt function and background level modelled using a polynomial function. The refined free parameters were background coefficients, scale factor, lattice constants, zero shift, peak profile, and asymmetry parameters.

For single BTO NC experiments, we prepared an aqueous colloidal suspension of BTO NC at the stock concentration of 1~mg/mL. 

\paragraph{BTO NC immobilization.} To enable contact mode PFM imaging on individual BTO NC, we immobilize them on a 170~$\mu$m thick glass coverslip with a conductive 80~nm-thick layer of indium-tin-oxide (ITO) deposited on it. Furthermore, a numbered grid made by gold-deposition and with the smallest cells of 30~$\mu$m size was  deposited on the ITO-coated coverslip. This allows us to address the same individual BTO NC after 90$^\circ$ rotation and also with other analytical methods. For immobilization while maintaining electrical contact, we first spin-coated a 40-80~nm thick layer of an aqueous suspension (at 1\% weight concentration) of a conductive polymer PEDOT:PSS (Ref.~768618, Sigma Aldrich) onto the marked ITO-coated coverslip. Subsequently, the stock solution of BTO NCs was first diluted by a factor of 50 in water, then spin-coated onto the coated coverslip and the sample was finally annealed at 95$^\circ$C for 30~minutes. While the literature suggests annealing PEDOT:PSS at temperatures ranging from 80$^\circ$C to 160$^\circ$C, we deliberately maintained the annealing temperature below 100$^\circ$C to avoid any potential phase transition of BTO, which typically occurs at a Curie temperature of 120$^\circ$C.

\subsection{PFM measurements}
For PFM we used a commercial apparatus (Dimension ICON head system and Nanoscope V controller, Bruker Inc.). The measurements were conducted using  SCM-PIT-V2 (Bruker) probes, which have conductive platinum-iridium coated tips with a radius of curvature of 25~nm and a nominal spring constant of 3~N/m. We acquired the piezoresponse phase and amplitude signal while scanning the tip along the laboratory frame $x$ axis direction (see Figure~\ref{fig:Experimental_config}b), with the cantilever along $y$ axis.

\paragraph{Minimization of electrostatic effects.}
It is widely recognized that PFM measurements can exhibit artifacts, leading to possible misinterpretations of the data. The signal artifacts are mainly caused by non-electromechanical effects such as electrostatic force that arises due to the electrostatic potential difference between the cantilever and the sample\cite{Balke.2015}. This force generates a bending response in the cantilever of the AFM that is not distinguishable from the true PFM signal. The strength of this force is highest near the tip-sample junction and decreases with increasing distance along the cantilever. To address this issue, we employed the concept of electrostatic blind spot position (ESBS) for the laser, as recently introduced by Killgore \textit{et al.} \cite{Killgore.2022}. In ESBS, the laser spot, which is usually placed just above the tip, is positioned along the cantilever where the bending induced by the electrostatic force is minimised, in order to get a piezoresponse amplitude dominated by the electromecanichal contribution. The signal thus obtained will be predominantly dominated by the true PFM signal caused by the bending of the cantilever due to inverse piezoelectric effect. The ESBS position was determined to be at a distance of 0.45 times the length of the cantilever.

PFM measurements were done on single isolated BTO NCs preidentified using scanning electron microscopy (SEM). The lateral PFM phase and amplitude images were taken using a drive voltage of 1.5~V and a drive frequency of 727.856 kHz (near in-plane contact resonance of the cantilever).
We used a scan speed of 0.35~Hz and utilized 192 scan lines across a full scan size of 300~nm resulting in 1.56~nm pixel size.

\paragraph{LPFM phase offset $\varphi_{\rm offset}$ estimation.}
To estimate $\varphi_{\rm offset}$, we first selected one domain with the smallest phase values, close to 0, as the reference, and computed a first phase offset as the phase average value within this domain. In parallel, we identified another domain with phase values close to 180$^\circ$. Subtracting 180$^\circ$ from the phase of this second domain led to a second phase offset. By averaging these offsets, we obtained a final offset value of $-12\pm\kern -1.8pt2$$^\circ$ for the PFM scan along the (O$X_{\rm c}$) direction. The same process was repeated for the PFM scan along the (O$Y_{\rm c}$) direction, producing an average offset value of $-9\pm\kern -1.8pt4$$^\circ$. As this offset is attributed to the instrumental response and should be the same for both scans, regardless of particle orientation, we took the average of these two values, resulting in $\varphi_{\rm offset}=-10\pm2^\circ$.

\paragraph{Vector lateral PFM mapping.} To build the $\boldsymbol{u}^{\rm m}$ lateral displacement field of probed-NC facets, we first had to register $u_{Y_{\rm c}}$ component map with $u_{X_{\rm c}}$ map, as they were acquired separately. We only considered translations shifts between the two scans, as rotation mismatch could not be assessed with a precision better than the uncertainty on the physical 90$^\circ$ sample rotation. The translation vector shift between $X_{\rm c}$ and $Y_{\rm c}$ components was determined using Fiji-ImageJ software~\cite{Schindelin.2012}, based on the largest overlap of the raw PFM amplitude images. We then represented the mixed displacement amplitude and vector map.

\subsection{Piezoresponse simulation}
Simulations of an embedded BTO nanoparticle have been performed using FERRET\cite{Mangeri2017}, a module for simulating multiferroic phenomena in the MOOSE finite element framework\cite{lindsay2022moose}.  
We consider the equilibrium domain structure and piezoresponse of a stress-free BTO nano-cube (100~nm size) embedded in a dielectric infinite medium (here assumed to be air), by evolving a time-dependent Landau-Ginzburg-Devonshire equation,
$\displaystyle \frac{\delta \boldsymbol{P}}{\delta t} =-\frac{\delta F}{\delta \boldsymbol{P}}$, with $F$ representing the free energy of the system. 

The Helmoltz free energy volume density $f$ includes several terms, introduced in Hlinka and Marton~\cite{Hlinka2006PhenomenologicalMO} and which analytical expressions are given in the Supporting Information Text~S2:
\begin{equation}
    f_{\rm H}\equiv f_{\rm bulk}+f_{\nabla{P}}+f_{\rm elastic}+f_{\rm electrostr}+f_{\rm elec},
    \label{eq:free_energy_density_sum}
\end{equation}
where $f_{\rm bulk}$ is the contribution to the energy stemming from the local polarization field $\boldsymbol{P}(\boldsymbol{r})$. In a uniaxial ferroelectric, this typically has two degenerate minima and a double-well shape as a function of the polarization value. This is typically the case in BTO and indicates the desire of the system to form a uniformly polarized ferroelectric domain. The term $f_{\nabla{P}}$ depends on the gradient of the polarization field and tends to oppose variations of the polarization field. The term $f_{\rm elastic}$ depends on the derivative of the local atomic displacement field $\boldsymbol{u}(\boldsymbol{r})$ (defined relative to a cubic parent lattice structure) and describes the linear elastic response of the system. The electrostrictive term $f_{\rm electrostr}$ gathers the electromechanical energy couplings between the polarization field $\boldsymbol{P}(\boldsymbol{r})$ and the displacement field $\boldsymbol{u}(\boldsymbol{r})$. Finally, $f_{\rm elec}$ represents the energy density caused by internal/external electric fields. 
Coupling to internal/external electrical fields is accomplished through the Poisson equation,
$\displaystyle \nabla^2\Phi= \frac{1}{\varepsilon_{\rm r}\varepsilon_0}\nabla.\boldsymbol{P}$, where $\Phi$ is the electric potential, related to the electric field $\boldsymbol{E}$ through $\boldsymbol{E}=-\nabla\Phi$, and $\varepsilon_{\rm r}$ is the surrounding medium relative permittivity, which in our case is air for most of the particle volume, hence $\varepsilon_{\rm r}=1$.
As stated earlier, as we consider a stress-free condition, the strain energy term $-\boldsymbol{\sigma.\varepsilon}$, where $\boldsymbol{\sigma}$ is the stress and $\boldsymbol{\varepsilon}$ the strain tensor, resumes at zero. 
To realize our simulations of ferroelectric domains and displacements on the BTO NC at the ambient temperature of 298~K, we adopted numerical coefficients from ref.~\cite{Hlinka2006PhenomenologicalMO} for the numerical values of the coefficients involved in each term of energy density.
The displacement field is obtained by solving the stress-divergence equation $\nabla.\boldsymbol{\sigma} = 0$ (neglecting gravity) with mechanical boundary conditions without stress at each time step.

The computational domain comprises the BTO nanocrystal,  with 100~nm long sides embedded at the center of a dielectric matrix,  with 200~nm sides in order to capture the depolarization field around the inclusion. 
The relative permittivity of the dielectric matrix is set to 1 and the NC is considered stress-free, simulating an open air environment. Dirichlet boundary conditions on ($u,\Phi$), are applied to ensure that displacement and electrostatic fields disappear far from the particle. Partial clamping and dielectric screening by PEDOT:PSS could lead to a slight alteration of the ground state and the piezoelectric response through modulation of the surface layer thickness and potential Poisson effects which could lead to enhanced out-of-plane mechanical response.

The equilibrium structure at zero applied electric field is obtained by relaxing the system starting from a reasonable approximation of the paraelectric state, realized by initializing random, small component ($<10^{-4}$C/m$^2$) values to the polarization and local displacement fields. Through the gradient flow approach, the system variables are evolved through a dynamic time stepper until a local minimum is reached, defined by a convergence criterion of within $10^{-6}\%$ of the total energy of the previous time step. 

The average thickness of the surface layers are estimated by the ratio of the volume having $P_{Z_{\rm c}} \approx 0$ to the volume of the nanocrystal, multiplied by the nanoparticle size. The obtained averaged thickness of 8~nm correlates well with typical thicknesses which can be estimated from Figure~\ref{fig:Polar_equilibrium_distribution}b.

To mimic PFM experiments, we use the following procedure. First, we start with the zero-field equilibrium structure. We then relax the polarization and local displacement fields by solving the Landau-Ginzburg-Devonshire equation and the stress-divergence equation at each time step, under an applied external electric field of 150~kV/cm, until the new equilibrium is reached. The electric field amplitude was chosen to closely match the one in the experiment of 120~kV/cm (corresponding to 1.5~V AC amplitude across a height of 125~nm for the NC displayed on Figure~\ref{fig:LPFM}). The difference in equilibrium local displacement fields between the configuration with and without applied electric field gives us the electric-field induced displacement in the BTO nanocube, which we report in Figure~\ref{fig:simul_domains_and_PFM_displacement}c,d and compare to lateral PFM experiments. Note that to compare with the experimental observations, a Gaussian blur with 20~nm diameter was applied to the local displacement field.

\section*{Author contributions}
Conceptualization: BD, CFD, CP, and FT; Methodology: AM, KC, KP, CFD, CP and FT; Software: KC and CP; Validation: AM, KC, CFD, CP and FT; Formal analysis: KC and CP; Investigation: AM, AZ, MV, and KC; Resources: CB and FK; Data curation: CB, MV, AM and KC; Writing – Original draft: AM, KC, CP and FT; Writing – Review and Editing: CFD, KP, CP and FT; Visualization: AM, KC, MV, AZ, KP, CP, and FT; Supervision: CFD and FT ; Project administration: CFD, CP and FT; Funding acquisition: BD, CP, CFD and FT.
All authors have read and agreed to the published version of the manuscript.


\begin{acknowledgement}
The authors thank Mathieu Chauvet for the fabrication of the control sample of $Y_{\rm c}$-cut PPLN and Pascale Gemeiner for the measurements of the temperature dependence of the BTO Raman spectrum.
The authors are grateful to Nataliya Alyabyeva for her initial support in PFM imaging and Dana Stanescu and Cindy Rountree for their training on the Bruker Icon PFM of the CEA/SPEC Interdisciplinary Multiscale Atomic Force Microscope Platform (IMAFMP). IMAFMP was supported by ``Triangle de la Physique'', ``PHOM'' and ``Ile-de-France'' Region (DIM-MAP, C’NanoIdF, and ISC-PIF). 
The authors thank Pierre Audebert, Guy Deniau, and Noelle Gogneau for their suggestions to bind the nanocrystals to the substrate; Laureen Moreau for her guidance in SEM imaging; and Simon Vassant for the realization of the marking grids enabling correlative SEM-PFM imaging. 
C.P. and K.C. acknowledge support from GENCI-TGCC computing resources through grant AD010913519 and  computational resources from the “Mésocentre” computing center of Université Paris-Saclay, CentraleSupélec and École Normale Supérieure Paris-Saclay supported by CNRS and Région Île-de-France (https://mesocentre.universite-paris-saclay.fr/).
This work has received financial support to B.D., C.F. and F.T. from the CNRS through the MITI interdisciplinary program and from the French National Research Agency (ANR, grant numbers ANR-21-CE09-0028 and ANR-21-CE09-0033).

\end{acknowledgement}

\section{Associated content}
The following preprint version of the work reported in this article is available: 
A. Muraleedharan, K.~Co, M.~Vallet, A.~Zaki, F.~Karolak, C.~Bogicevic, K.~Perronet, B.~Dkhil, C.~Paillard, C.~Fiorini-Debbuischert, and F.~Treussart. Ferroelectric texture of individual barium titanate nanocrystals. \textbf{2024}, 2402.14502 [cond-mat.mtrl-sci]. \underline{arXiv}. \url{https://arxiv.org/abs/2402.14502} (accessed June 13, 2024).
\begin{suppinfo}
The following file is available free of charge.

Supporting information file comprising:
\begin{itemize}
	\item Two supporting texts: PFM basics; analytical expressions of the contributions to the free energy volume density used in phase field simulations.
	\item Supporting data figures: Additional characterizations of the BTO sample (morphology and strain, HR-STEM images, temperature-dependent Raman spectroscopy and Curie temperature determination, PEDOT:PSS layer thickness measurement, size distribution of BTO NC inferred from SEM measurements); Evidence of non-electromechanical effects and use of ESBS setting to suppress them; LPFM resonance curve of a cantilever; Lateral and vertical PFM measurements on a reference sample with domains that can be either oriented in-plane or out-of-plane; Lateral PFM mapping of three additional BTO NC; Additional data on phase field simulations (dielectric constant as a function of the nanocrystal size, distribution of polarization in facets orthogonal or parallel to the polar axis, distribution of the different components of the energy density across a NC).
\end{itemize}
\end{suppinfo}


\providecommand{\latin}[1]{#1}
\makeatletter
\providecommand{\doi}
{\begingroup\let\do\@makeother\dospecials
	\catcode`\{=1 \catcode`\}=2 \doi@aux}
\providecommand{\doi@aux}[1]{\endgroup\texttt{#1}}
\makeatother
\providecommand*\mcitethebibliography{\thebibliography}
\csname @ifundefined\endcsname{endmcitethebibliography}
{\let\endmcitethebibliography\endthebibliography}{}


\begin{mcitethebibliography}{72}
	\providecommand*\natexlab[1]{#1}
	\providecommand*\mciteSetBstSublistMode[1]{}
	\providecommand*\mciteSetBstMaxWidthForm[2]{}
	\providecommand*\mciteBstWouldAddEndPuncttrue
	{\def\EndOfBibitem{\unskip.}}
	\providecommand*\mciteBstWouldAddEndPunctfalse
	{\let\EndOfBibitem\relax}
	\providecommand*\mciteSetBstMidEndSepPunct[3]{}
	\providecommand*\mciteSetBstSublistLabelBeginEnd[3]{}
	\providecommand*\EndOfBibitem{}
	\mciteSetBstSublistMode{f}
	\mciteSetBstMaxWidthForm{subitem}{(\alph{mcitesubitemcount})}
	\mciteSetBstSublistLabelBeginEnd
	{\mcitemaxwidthsubitemform\space}
	{\relax}
	{\relax}
	
	\bibitem[Mikolajick \latin{et~al.}(2021)Mikolajick, Slesazeck, Mulaosmanovic,
	Park, Fichtner, Lomenzo, Hoffmann, and Schroeder]{Mikolajick2021}
	Mikolajick,~T.; Slesazeck,~S.; Mulaosmanovic,~H.; Park,~M.~H.; Fichtner,~S.;
	Lomenzo,~P.~D.; Hoffmann,~M.; Schroeder,~U. {Next generation ferroelectric
		materials for semiconductor process integration and their applications}.
	\emph{Journal of Applied Physics} \textbf{2021}, \emph{129}\relax
	\mciteBstWouldAddEndPuncttrue
	\mciteSetBstMidEndSepPunct{\mcitedefaultmidpunct}
	{\mcitedefaultendpunct}{\mcitedefaultseppunct}\relax
	\EndOfBibitem
	\bibitem[Naumov \latin{et~al.}(2004)Naumov, Bellaiche, and Fu]{Naumov2004}
	Naumov,~I.~I.; Bellaiche,~L.; Fu,~H. {Unusual phase transitions in
		ferroelectric nanodisks and nanorods}. \emph{Nature} \textbf{2004},
	\emph{432}, 737--740\relax
	\mciteBstWouldAddEndPuncttrue
	\mciteSetBstMidEndSepPunct{\mcitedefaultmidpunct}
	{\mcitedefaultendpunct}{\mcitedefaultseppunct}\relax
	\EndOfBibitem
	\bibitem[Louis \latin{et~al.}(2012)Louis, Kornev, Geneste, Dkhil, and
	Bellaiche]{Louis2012}
	Louis,~L.; Kornev,~I.~A.; Geneste,~G.; Dkhil,~B.; Bellaiche,~L. {Novel complex
		phenomena in ferroelectric nanocomposites}. \emph{Journal of Physics:
		Condensed Matter} \textbf{2012}, \emph{24}, 402201\relax
	\mciteBstWouldAddEndPuncttrue
	\mciteSetBstMidEndSepPunct{\mcitedefaultmidpunct}
	{\mcitedefaultendpunct}{\mcitedefaultseppunct}\relax
	\EndOfBibitem
	\bibitem[Prosandeev and Bellaiche(2008)Prosandeev, and
	Bellaiche]{Prosandeev2008a}
	Prosandeev,~S.; Bellaiche,~L. {Controlling Double Vortex States in
		Low-Dimensional Dipolar Systems}. \emph{Physical Review Letters}
	\textbf{2008}, \emph{101}, 097203\relax
	\mciteBstWouldAddEndPuncttrue
	\mciteSetBstMidEndSepPunct{\mcitedefaultmidpunct}
	{\mcitedefaultendpunct}{\mcitedefaultseppunct}\relax
	\EndOfBibitem
	\bibitem[Nahas \latin{et~al.}(2015)Nahas, Prokhorenko, Louis, Gui, Kornev, and
	Bellaiche]{Nahas2015}
	Nahas,~Y.; Prokhorenko,~S.; Louis,~L.; Gui,~Z.; Kornev,~I.~A.; Bellaiche,~L.
	{Discovery of stable skyrmionic state in ferroelectric nanocomposites}.
	\emph{Nature Communications} \textbf{2015}, \emph{6}, 8542\relax
	\mciteBstWouldAddEndPuncttrue
	\mciteSetBstMidEndSepPunct{\mcitedefaultmidpunct}
	{\mcitedefaultendpunct}{\mcitedefaultseppunct}\relax
	\EndOfBibitem
	\bibitem[Nahas \latin{et~al.}(2017)Nahas, Prokhorenko, Kornev, and
	Bellaiche]{Nahas2017a}
	Nahas,~Y.; Prokhorenko,~S.; Kornev,~I.~A.; Bellaiche,~L. {Emergent
		Berezinskii-Kosterlitz-Thouless Phase in Low-Dimensional Ferroelectrics}.
	\emph{Physical Review Letters} \textbf{2017}, \emph{119}, 117601\relax
	\mciteBstWouldAddEndPuncttrue
	\mciteSetBstMidEndSepPunct{\mcitedefaultmidpunct}
	{\mcitedefaultendpunct}{\mcitedefaultseppunct}\relax
	\EndOfBibitem
	\bibitem[Gon{\c{c}}alves \latin{et~al.}(2019)Gon{\c{c}}alves,
	Escorihuela-Sayalero, Garca-Fern{\'{a}}ndez, Junquera, and
	{\'{I}}{\~{n}}iguez]{Goncalves2019}
	Gon{\c{c}}alves,~M.~A.; Escorihuela-Sayalero,~C.; Garca-Fern{\'{a}}ndez,~P.;
	Junquera,~J.; {\'{I}}{\~{n}}iguez,~J. {Theoretical guidelines to create and
		tune electric skyrmion bubbles}. \emph{Science Advances} \textbf{2019},
	\emph{5}, 1--6\relax
	\mciteBstWouldAddEndPuncttrue
	\mciteSetBstMidEndSepPunct{\mcitedefaultmidpunct}
	{\mcitedefaultendpunct}{\mcitedefaultseppunct}\relax
	\EndOfBibitem
	\bibitem[Balke \latin{et~al.}(2011)Balke, Winchester, Ren, Chu, Morozovska,
	Eliseev, Huijben, Vasudevan, Maksymovych, Britson, Jesse, Kornev, Ramesh,
	Bellaiche, Chen, and Kalinin]{Balke2011}
	Balke,~N.; Winchester,~B.; Ren,~W.; Chu,~Y.~H.; Morozovska,~A.~N.;
	Eliseev,~E.~a.; Huijben,~M.; Vasudevan,~R.~K.; Maksymovych,~P.; Britson,~J.;
	Jesse,~S.; Kornev,~I.~A.; Ramesh,~R.; Bellaiche,~L.; Chen,~L.-Q.;
	Kalinin,~S.~V. {Enhanced electric conductivity at ferroelectric vortex cores
		in BiFeO$_3$}. \emph{Nature Physics} \textbf{2011}, \emph{8}, 81--88\relax
	\mciteBstWouldAddEndPuncttrue
	\mciteSetBstMidEndSepPunct{\mcitedefaultmidpunct}
	{\mcitedefaultendpunct}{\mcitedefaultseppunct}\relax
	\EndOfBibitem
	\bibitem[Yadav \latin{et~al.}(2016)Yadav, Nelson, Hsu, Hong, Clarkson,
	Schlep{\"{u}}tz, Damodaran, Shafer, Arenholz, Dedon, Chen, Vishwanath, Minor,
	Chen, Scott, Martin, and Ramesh]{Yadav2016}
	Yadav,~A.~K.; Nelson,~C.~T.; Hsu,~S.~L.; Hong,~Z.; Clarkson,~J.~D.;
	Schlep{\"{u}}tz,~C.~M.; Damodaran,~A.~R.; Shafer,~P.; Arenholz,~E.;
	Dedon,~L.~R.; Chen,~D.; Vishwanath,~A.; Minor,~A.~M.; Chen,~L.-Q.;
	Scott,~J.~F.; Martin,~L.~W.; Ramesh,~R. {Observation of polar vortices in
		oxide superlattices}. \emph{Nature} \textbf{2016}, \emph{530}, 198--201\relax
	\mciteBstWouldAddEndPuncttrue
	\mciteSetBstMidEndSepPunct{\mcitedefaultmidpunct}
	{\mcitedefaultendpunct}{\mcitedefaultseppunct}\relax
	\EndOfBibitem
	\bibitem[Das \latin{et~al.}(2019)Das, Tang, Hong, Gon{\c{c}}alves, McCarter,
	Klewe, Nguyen, G{\'{o}}mez-Ortiz, Shafer, Arenholz, Stoica, Hsu, Wang, Ophus,
	Liu, Nelson, Saremi, Prasad, Mei, Schlom, {\'{I}}{\~{n}}iguez,
	Garc{\'{i}}a-Fern{\'{a}}ndez, Muller, Chen, Junquera, Martin, and
	Ramesh]{Das2019}
	Das,~S.; Tang,~Y.~L.; Hong,~Z.; Gon{\c{c}}alves,~M. A.~P.; McCarter,~M.~R.;
	Klewe,~C.; Nguyen,~K.~X.; G{\'{o}}mez-Ortiz,~F.; Shafer,~P.; Arenholz,~E.;
	Stoica,~V.~A.; Hsu,~S.-L.; Wang,~B.; Ophus,~C.; Liu,~J.~F.; Nelson,~C.~T.;
	Saremi,~S.; Prasad,~B.; Mei,~A.~B.; Schlom,~D.~G. \latin{et~al.}
	{Observation of room-temperature polar skyrmions}. \emph{Nature}
	\textbf{2019}, \emph{568}, 368--372\relax
	\mciteBstWouldAddEndPuncttrue
	\mciteSetBstMidEndSepPunct{\mcitedefaultmidpunct}
	{\mcitedefaultendpunct}{\mcitedefaultseppunct}\relax
	\EndOfBibitem
	\bibitem[Abid \latin{et~al.}(2021)Abid, Sun, Hou, Tan, Zhong, Zhu, Chen, Qu,
	Li, Wu, Zhang, Wang, Liu, Bai, Yu, Ouyang, Wang, Li, and Gao]{Abid2021}
	Abid,~A.~Y.; Sun,~Y.; Hou,~X.; Tan,~C.; Zhong,~X.; Zhu,~R.; Chen,~H.; Qu,~K.;
	Li,~Y.; Wu,~M.; Zhang,~J.; Wang,~J.; Liu,~K.; Bai,~X.; Yu,~D.; Ouyang,~X.;
	Wang,~J.; Li,~J.; Gao,~P. {Creating polar antivortex in PbTiO$_3$/SrTiO$_3$
		superlattice}. \emph{Nature Communications} \textbf{2021}, \emph{12},
	2054\relax
	\mciteBstWouldAddEndPuncttrue
	\mciteSetBstMidEndSepPunct{\mcitedefaultmidpunct}
	{\mcitedefaultendpunct}{\mcitedefaultseppunct}\relax
	\EndOfBibitem
	\bibitem[Tan \latin{et~al.}(2021)Tan, Dong, Sun, Liu, Chen, Zhong, Zhu, Liu,
	Zhang, Wang, Liu, Bai, Yu, Ouyang, Wang, Gao, Luo, and Li]{Tan2021}
	Tan,~C.; Dong,~Y.; Sun,~Y.; Liu,~C.; Chen,~P.; Zhong,~X.; Zhu,~R.; Liu,~M.;
	Zhang,~J.; Wang,~J.; Liu,~K.; Bai,~X.; Yu,~D.; Ouyang,~X.; Wang,~J.; Gao,~P.;
	Luo,~Z.; Li,~J. {Engineering polar vortex from topologically trivial domain
		architecture}. \emph{Nature Communications} \textbf{2021}, \emph{12},
	4620\relax
	\mciteBstWouldAddEndPuncttrue
	\mciteSetBstMidEndSepPunct{\mcitedefaultmidpunct}
	{\mcitedefaultendpunct}{\mcitedefaultseppunct}\relax
	\EndOfBibitem
	\bibitem[Das \latin{et~al.}(2021)Das, Hong, Stoica, Gon{\c{c}}alves, Shao,
	Parsonnet, Marksz, Saremi, McCarter, Reynoso, Long, Hagerstrom, Meyers, Ravi,
	Prasad, Zhou, Zhang, Wen, G{\'{o}}mez-Ortiz, Garc{\'{i}}a-Fern{\'{a}}ndez,
	Bokor, {\'{I}}{\~{n}}iguez, Freeland, Orloff, Junquera, Chen, Salahuddin,
	Muller, Martin, and Ramesh]{Das2021}
	Das,~S.; Hong,~Z.; Stoica,~V.~A.; Gon{\c{c}}alves,~M. A.~P.; Shao,~Y.~T.;
	Parsonnet,~E.; Marksz,~E.~J.; Saremi,~S.; McCarter,~M.~R.; Reynoso,~A.;
	Long,~C.~J.; Hagerstrom,~A.~M.; Meyers,~D.; Ravi,~V.; Prasad,~B.; Zhou,~H.;
	Zhang,~Z.; Wen,~H.; G{\'{o}}mez-Ortiz,~F.; Garc{\'{i}}a-Fern{\'{a}}ndez,~P.
	\latin{et~al.}  {Local negative permittivity and topological phase transition
		in polar skyrmions}. \emph{Nature Materials} \textbf{2021}, \emph{20},
	194--201\relax
	\mciteBstWouldAddEndPuncttrue
	\mciteSetBstMidEndSepPunct{\mcitedefaultmidpunct}
	{\mcitedefaultendpunct}{\mcitedefaultseppunct}\relax
	\EndOfBibitem
	\bibitem[Li \latin{et~al.}(2017)Li, Wang, Tian, Li, Zhao, Zhang, Yao, Fan,
	Song, Chen, Fan, Qin, Zeng, Zhang, Lu, Hu, Lei, Zhu, Li, Gao, and
	Liu]{Li2017}
	Li,~Z.; Wang,~Y.; Tian,~G.; Li,~P.; Zhao,~L.; Zhang,~F.; Yao,~J.; Fan,~H.;
	Song,~X.; Chen,~D.; Fan,~Z.; Qin,~M.; Zeng,~M.; Zhang,~Z.; Lu,~X.; Hu,~S.;
	Lei,~C.; Zhu,~Q.; Li,~J.; Gao,~X. \latin{et~al.}  {High-density array of
		ferroelectric nanodots with robust and reversibly switchable topological
		domain states}. \emph{Science Advances} \textbf{2017}, \emph{3}, 1--9\relax
	\mciteBstWouldAddEndPuncttrue
	\mciteSetBstMidEndSepPunct{\mcitedefaultmidpunct}
	{\mcitedefaultendpunct}{\mcitedefaultseppunct}\relax
	\EndOfBibitem
	\bibitem[Mangeri \latin{et~al.}(2017)Mangeri, Espinal, Jokisaari, {Pamir
		Alpay}, Nakhmanson, and Heinonen]{Mangeri2017}
	Mangeri,~J.; Espinal,~Y.; Jokisaari,~A.; {Pamir Alpay},~S.; Nakhmanson,~S.;
	Heinonen,~O. {Topological phase transformations and intrinsic size effects in
		ferroelectric nanoparticles}. \emph{Nanoscale} \textbf{2017}, \emph{9},
	1616--1624\relax
	\mciteBstWouldAddEndPuncttrue
	\mciteSetBstMidEndSepPunct{\mcitedefaultmidpunct}
	{\mcitedefaultendpunct}{\mcitedefaultseppunct}\relax
	\EndOfBibitem
	\bibitem[Luk'yanchuk \latin{et~al.}(2020)Luk'yanchuk, Tikhonov, Razumnaya, and
	Vinokur]{Lukyanchuk2020}
	Luk'yanchuk,~I.; Tikhonov,~Y.; Razumnaya,~A.; Vinokur,~V.~M. {Hopfions emerge
		in ferroelectrics}. \emph{Nature Communications} \textbf{2020}, \emph{11},
	2433\relax
	\mciteBstWouldAddEndPuncttrue
	\mciteSetBstMidEndSepPunct{\mcitedefaultmidpunct}
	{\mcitedefaultendpunct}{\mcitedefaultseppunct}\relax
	\EndOfBibitem
	\bibitem[Co \latin{et~al.}(2021)Co, {Pamir Alpay}, Nakhmanson, and
	Mangeri]{Co2021}
	Co,~K.; {Pamir Alpay},~S.; Nakhmanson,~S.; Mangeri,~J. {Surface charge mediated
		polar response in ferroelectric nanoparticles}. \emph{Applied Physics
		Letters} \textbf{2021}, \emph{119}\relax
	\mciteBstWouldAddEndPuncttrue
	\mciteSetBstMidEndSepPunct{\mcitedefaultmidpunct}
	{\mcitedefaultendpunct}{\mcitedefaultseppunct}\relax
	\EndOfBibitem
	\bibitem[Kwei \latin{et~al.}(1993)Kwei, Lawson, Billinge, and
	Cheong]{Kwei.1993}
	Kwei,~G.~H.; Lawson,~A.~C.; Billinge,~S. J.~L.; Cheong,~S.~W. {Structures of
		the ferroelectric phases of barium titanate}. \emph{The Journal of Physical
		Chemistry} \textbf{1993}, \emph{97}, 2368--2377\relax
	\mciteBstWouldAddEndPuncttrue
	\mciteSetBstMidEndSepPunct{\mcitedefaultmidpunct}
	{\mcitedefaultendpunct}{\mcitedefaultseppunct}\relax
	\EndOfBibitem
	\bibitem[Bogicevic \latin{et~al.}(2015)Bogicevic, Thorner, Karolak,
	Haghi-Ashtiani, and Kiat]{Bogicevic.2015}
	Bogicevic,~C.; Thorner,~G.; Karolak,~F.; Haghi-Ashtiani,~P.; Kiat,~J.-M.
	{Morphogenesis mechanisms in the solvothermal synthesis of BaTiO$_3$ from
		titanate nanorods and nanotubes}. \emph{Nanoscale} \textbf{2015}, \emph{7},
	3594--3603\relax
	\mciteBstWouldAddEndPuncttrue
	\mciteSetBstMidEndSepPunct{\mcitedefaultmidpunct}
	{\mcitedefaultendpunct}{\mcitedefaultseppunct}\relax
	\EndOfBibitem
	\bibitem[Jiang \latin{et~al.}(2019)Jiang, Iocozzia, Zhao, Zhang, Harn, Chen,
	and Lin]{Jiang.2019}
	Jiang,~B.; Iocozzia,~J.; Zhao,~L.; Zhang,~H.; Harn,~Y.-W.; Chen,~Y.; Lin,~Z.
	{Barium titanate at the nanoscale: controlled synthesis and dielectric and
		ferroelectric properties}. \emph{Chemical Society Reviews} \textbf{2019},
	\emph{48}, 1194--1228\relax
	\mciteBstWouldAddEndPuncttrue
	\mciteSetBstMidEndSepPunct{\mcitedefaultmidpunct}
	{\mcitedefaultendpunct}{\mcitedefaultseppunct}\relax
	\EndOfBibitem
	\bibitem[Smith \latin{et~al.}(2008)Smith, Page, Siegrist, Redmond, Walter,
	Seshadri, Brus, and Steigerwald]{Smith.2008}
	Smith,~M.~B.; Page,~K.; Siegrist,~T.; Redmond,~P.~L.; Walter,~E.~C.;
	Seshadri,~R.; Brus,~L.~E.; Steigerwald,~M.~L. {Crystal Structure and the
		Paraelectric-to-Ferroelectric Phase Transition of Nanoscale BaTiO$_3$}.
	\emph{Journal of the American Chemical Society} \textbf{2008}, \emph{130},
	6955--6963\relax
	\mciteBstWouldAddEndPuncttrue
	\mciteSetBstMidEndSepPunct{\mcitedefaultmidpunct}
	{\mcitedefaultendpunct}{\mcitedefaultseppunct}\relax
	\EndOfBibitem
	\bibitem[Cui \latin{et~al.}(2013)Cui, Briscoe, and Dunn]{Cui2013}
	Cui,~Y.; Briscoe,~J.; Dunn,~S. {Effect of Ferroelectricity on
		Solar-Light-Driven Photocatalytic Activity of BaTiO$_3$ —Influence on the
		Carrier Separation and Stern Layer Formation}. \emph{Chemistry of Materials}
	\textbf{2013}, \emph{25}, 4215--4223\relax
	\mciteBstWouldAddEndPuncttrue
	\mciteSetBstMidEndSepPunct{\mcitedefaultmidpunct}
	{\mcitedefaultendpunct}{\mcitedefaultseppunct}\relax
	\EndOfBibitem
	\bibitem[Paillard \latin{et~al.}(2016)Paillard, Bai, Infante, Guennou, Geneste,
	Alexe, Kreisel, and Dkhil]{Paillard2016}
	Paillard,~C.; Bai,~X.; Infante,~I.~C.; Guennou,~M.; Geneste,~G.; Alexe,~M.;
	Kreisel,~J.; Dkhil,~B. {Photovoltaics with Ferroelectrics: Current Status and
		Beyond}. \emph{Advanced Materials} \textbf{2016}, \emph{28}, 5153--5168\relax
	\mciteBstWouldAddEndPuncttrue
	\mciteSetBstMidEndSepPunct{\mcitedefaultmidpunct}
	{\mcitedefaultendpunct}{\mcitedefaultseppunct}\relax
	\EndOfBibitem
	\bibitem[Li \latin{et~al.}(2020)Li, Li, Yang, and Wang]{Li2020a}
	Li,~Y.; Li,~J.; Yang,~W.; Wang,~X. {Implementation of ferroelectric materials
		in photocatalytic and photoelectrochemical water splitting}. \emph{Nanoscale
		Horizons} \textbf{2020}, \emph{5}, 1174--1187\relax
	\mciteBstWouldAddEndPuncttrue
	\mciteSetBstMidEndSepPunct{\mcitedefaultmidpunct}
	{\mcitedefaultendpunct}{\mcitedefaultseppunct}\relax
	\EndOfBibitem
	\bibitem[Hao \latin{et~al.}(2021)Hao, Feng, Banerjee, Wang, Billinge, Wang,
	Jin, Bi, and Li]{Hao2021}
	Hao,~Y.; Feng,~Z.; Banerjee,~S.; Wang,~X.; Billinge,~S.~J.; Wang,~J.; Jin,~K.;
	Bi,~K.; Li,~L. {Ferroelectric state and polarization switching behaviour of
		ultrafine BaTiO$_3$ nanoparticles with large-scale size uniformity}.
	\emph{Journal of Materials Chemistry C} \textbf{2021}, \emph{9},
	5267--5276\relax
	\mciteBstWouldAddEndPuncttrue
	\mciteSetBstMidEndSepPunct{\mcitedefaultmidpunct}
	{\mcitedefaultendpunct}{\mcitedefaultseppunct}\relax
	\EndOfBibitem
	\bibitem[Abbasi \latin{et~al.}(2022)Abbasi, Barone, Cruz-Jáuregui,
	Valdespino-Padilla, Paik, Kim, Kornblum, Schlom, Pascal, and
	Fenning]{Abbasi.2022}
	Abbasi,~P.; Barone,~M.~R.; Cruz-Jáuregui,~M. d. l.~P.; Valdespino-Padilla,~D.;
	Paik,~H.; Kim,~T.; Kornblum,~L.; Schlom,~D.~G.; Pascal,~T.~A.; Fenning,~D.~P.
	{Ferroelectric Modulation of Surface Electronic States in BaTiO$_3$ for
		Enhanced Hydrogen Evolution Activity}. \emph{Nano Letters} \textbf{2022},
	\emph{22}, 4276--4284\relax
	\mciteBstWouldAddEndPuncttrue
	\mciteSetBstMidEndSepPunct{\mcitedefaultmidpunct}
	{\mcitedefaultendpunct}{\mcitedefaultseppunct}\relax
	\EndOfBibitem
	\bibitem[Neige \latin{et~al.}(2023)Neige, Schwab, Musso, Berger, Bourret, and
	Diwald]{Neige.2023}
	Neige,~E.; Schwab,~T.; Musso,~M.; Berger,~T.; Bourret,~G.~R.; Diwald,~O.
	{Charge Separation in BaTiO$_3$ Nanocrystals: Spontaneous Polarization Versus
		Point Defect Chemistry}. \emph{Small} \textbf{2023}, \emph{19},
	e2206805\relax
	\mciteBstWouldAddEndPuncttrue
	\mciteSetBstMidEndSepPunct{\mcitedefaultmidpunct}
	{\mcitedefaultendpunct}{\mcitedefaultseppunct}\relax
	\EndOfBibitem
	\bibitem[Assavachin and Osterloh(2023)Assavachin, and
	Osterloh]{Assavachin.2023}
	Assavachin,~S.; Osterloh,~F.~E. {Ferroelectric Polarization in BaTiO$_3$
		Nanocrystals Controls Photoelectrochemical Water Oxidation and Photocatalytic
		Hydrogen Evolution}. \emph{Journal of the American Chemical Society}
	\textbf{2023}, \emph{145}, 18825--18833\relax
	\mciteBstWouldAddEndPuncttrue
	\mciteSetBstMidEndSepPunct{\mcitedefaultmidpunct}
	{\mcitedefaultendpunct}{\mcitedefaultseppunct}\relax
	\EndOfBibitem
	\bibitem[Schilling \latin{et~al.}(2009)Schilling, Byrne, Catalan, Webber,
	Genenko, Wu, Scott, and Gregg]{Schilling.2009}
	Schilling,~A.; Byrne,~D.; Catalan,~G.; Webber,~K.~G.; Genenko,~Y.~A.;
	Wu,~G.~S.; Scott,~J.~F.; Gregg,~J.~M. {Domains in Ferroelectric Nanodots}.
	\emph{Nano Letters} \textbf{2009}, \emph{9}, 3359--3364\relax
	\mciteBstWouldAddEndPuncttrue
	\mciteSetBstMidEndSepPunct{\mcitedefaultmidpunct}
	{\mcitedefaultendpunct}{\mcitedefaultseppunct}\relax
	\EndOfBibitem
	\bibitem[Denneulin and Everhardt(2022)Denneulin, and Everhardt]{Denneulin.2022}
	Denneulin,~T.; Everhardt,~A.~S. {A transmission electron microscopy study of
		low-strain epitaxial BaTiO$_3$ grown onto NdScO$_3$}. \emph{Journal of
		Physics: Condensed Matter} \textbf{2022}, \emph{34}, 235701\relax
	\mciteBstWouldAddEndPuncttrue
	\mciteSetBstMidEndSepPunct{\mcitedefaultmidpunct}
	{\mcitedefaultendpunct}{\mcitedefaultseppunct}\relax
	\EndOfBibitem
	\bibitem[Campanini \latin{et~al.}(2019)Campanini, Erni, and
	Rossell]{Campanini.2019}
	Campanini,~M.; Erni,~R.; Rossell,~M.~D. {Probing local order in multiferroics
		by transmission electron microscopy}. \emph{Physical Sciences Reviews}
	\textbf{2019}, \emph{5}, 20190068\relax
	\mciteBstWouldAddEndPuncttrue
	\mciteSetBstMidEndSepPunct{\mcitedefaultmidpunct}
	{\mcitedefaultendpunct}{\mcitedefaultseppunct}\relax
	\EndOfBibitem
	\bibitem[Polking \latin{et~al.}(2012)Polking, Han, Yourdkhani, Petkov,
	Kisielowski, Volkov, Zhu, Caruntu, Alivisatos, and Ramesh]{Polking.2012}
	Polking,~M.~J.; Han,~M.-G.; Yourdkhani,~A.; Petkov,~V.; Kisielowski,~C.~F.;
	Volkov,~V.~V.; Zhu,~Y.; Caruntu,~G.; Alivisatos,~A.~P.; Ramesh,~R.
	{Ferroelectric order in individual nanometre-scale crystals}. \emph{Nature
		Materials} \textbf{2012}, \emph{11}, 700--709\relax
	\mciteBstWouldAddEndPuncttrue
	\mciteSetBstMidEndSepPunct{\mcitedefaultmidpunct}
	{\mcitedefaultendpunct}{\mcitedefaultseppunct}\relax
	\EndOfBibitem
	\bibitem[Karpov \latin{et~al.}(2017)Karpov, Liu, Rolo, Harder, Balachandran,
	Xue, Lookman, and Fohtung]{Karpov.2017}
	Karpov,~D.; Liu,~Z.; Rolo,~T. d.~S.; Harder,~R.; Balachandran,~P.~V.; Xue,~D.;
	Lookman,~T.; Fohtung,~E. {Three-dimensional imaging of vortex structure in a
		ferroelectric nanoparticle driven by an electric field}. \emph{Nature
		Communications} \textbf{2017}, \emph{8}, 280\relax
	\mciteBstWouldAddEndPuncttrue
	\mciteSetBstMidEndSepPunct{\mcitedefaultmidpunct}
	{\mcitedefaultendpunct}{\mcitedefaultseppunct}\relax
	\EndOfBibitem
	\bibitem[Megaw(1945)]{Megaw.1945}
	Megaw,~H. {Crystal Structure of Barium Titanate}. \emph{Nature} \textbf{1945},
	\emph{155}, 484--485\relax
	\mciteBstWouldAddEndPuncttrue
	\mciteSetBstMidEndSepPunct{\mcitedefaultmidpunct}
	{\mcitedefaultendpunct}{\mcitedefaultseppunct}\relax
	\EndOfBibitem
	\bibitem[Huang \latin{et~al.}(2017)Huang, Lu, Li, Tang, Yao, Tao, Liang, and
	Lu]{Huang.2017ar}
	Huang,~Y.; Lu,~B.; Li,~D.; Tang,~Z.; Yao,~Y.; Tao,~T.; Liang,~B.; Lu,~S.
	{Control of tetragonality via dehydroxylation of BaTiO$_3$ ultrafine
		powders}. \emph{Ceramics International} \textbf{2017}, \emph{43},
	16462--16466\relax
	\mciteBstWouldAddEndPuncttrue
	\mciteSetBstMidEndSepPunct{\mcitedefaultmidpunct}
	{\mcitedefaultendpunct}{\mcitedefaultseppunct}\relax
	\EndOfBibitem
	\bibitem[Lee \latin{et~al.}(2012)Lee, Moon, Choi, and Kim]{Lee.2012}
	Lee,~H.; Moon,~S.; Choi,~C.; Kim,~D.~K. {Synthesis and Size Control of
		Tetragonal Barium Titanate Nanopowders by Facile Solvothermal Method}.
	\emph{Journal of the American Ceramic Society} \textbf{2012}, \emph{95},
	2429--2434\relax
	\mciteBstWouldAddEndPuncttrue
	\mciteSetBstMidEndSepPunct{\mcitedefaultmidpunct}
	{\mcitedefaultendpunct}{\mcitedefaultseppunct}\relax
	\EndOfBibitem
	\bibitem[Zhu \latin{et~al.}(2009)Zhu, Zhang, Zhu, Zhou, and Liu]{Zhu.2009}
	Zhu,~X.; Zhang,~Z.; Zhu,~J.; Zhou,~S.; Liu,~Z. {Morphology and atomic-scale
		surface structure of barium titanate nanocrystals formed at hydrothermal
		conditions}. \emph{Journal of Crystal Growth} \textbf{2009}, \emph{311},
	2437--2442\relax
	\mciteBstWouldAddEndPuncttrue
	\mciteSetBstMidEndSepPunct{\mcitedefaultmidpunct}
	{\mcitedefaultendpunct}{\mcitedefaultseppunct}\relax
	\EndOfBibitem
	\bibitem[Vivekanandan and Kutty(1989)Vivekanandan, and
	Kutty]{Vivekanandan.1989}
	Vivekanandan,~R.; Kutty,~T. {Characterization of barium titanate fine powders
		formed from hydrothermal crystallization}. \emph{Powder Technology}
	\textbf{1989}, \emph{57}, 181--192\relax
	\mciteBstWouldAddEndPuncttrue
	\mciteSetBstMidEndSepPunct{\mcitedefaultmidpunct}
	{\mcitedefaultendpunct}{\mcitedefaultseppunct}\relax
	\EndOfBibitem
	\bibitem[Suzana \latin{et~al.}(2023)Suzana, Liu, Diao, Wu, Assefa, Abeykoon,
	Harder, Cha, Bozin, and Robinson]{Suzana.2023}
	Suzana,~A.~F.; Liu,~S.; Diao,~J.; Wu,~L.; Assefa,~T.~A.; Abeykoon,~M.;
	Harder,~R.; Cha,~W.; Bozin,~E.~S.; Robinson,~I.~K. {Structural Explanation of
		the Dielectric Enhancement of Barium Titanate Nanoparticles Grown under
		Hydrothermal Conditions}. \emph{Advanced Functional Materials} \textbf{2023},
	\emph{33}, 2208012\relax
	\mciteBstWouldAddEndPuncttrue
	\mciteSetBstMidEndSepPunct{\mcitedefaultmidpunct}
	{\mcitedefaultendpunct}{\mcitedefaultseppunct}\relax
	\EndOfBibitem
	\bibitem[DiDomenico \latin{et~al.}(1968)DiDomenico, Wemple, Porto, and
	Bauman]{DiDomenico.1968}
	DiDomenico,~M.; Wemple,~S.~H.; Porto,~S. P.~S.; Bauman,~R.~P. {Raman Spectrum
		of Single-Domain BaTiO$_3$}. \emph{Physical Review} \textbf{1968},
	\emph{174}, 522--530\relax
	\mciteBstWouldAddEndPuncttrue
	\mciteSetBstMidEndSepPunct{\mcitedefaultmidpunct}
	{\mcitedefaultendpunct}{\mcitedefaultseppunct}\relax
	\EndOfBibitem
	\bibitem[Begg \latin{et~al.}(1996)Begg, Finnie, and Vance]{Begg.1996}
	Begg,~B.~D.; Finnie,~K.~S.; Vance,~E.~R. {Raman Study of the Relationship
		between Room‐Temperature Tetragonality and the Curie Point of Barium
		Titanate}. \emph{Journal of the American Ceramic Society} \textbf{1996},
	\emph{79}, 2666--2672\relax
	\mciteBstWouldAddEndPuncttrue
	\mciteSetBstMidEndSepPunct{\mcitedefaultmidpunct}
	{\mcitedefaultendpunct}{\mcitedefaultseppunct}\relax
	\EndOfBibitem
	\bibitem[Kalinin \latin{et~al.}(2006)Kalinin, Rodriguez, Jesse, Shin, Baddorf,
	Gupta, Jain, Williams, and Gruverman]{Kalinin.2006}
	Kalinin,~S.~V.; Rodriguez,~B.~J.; Jesse,~S.; Shin,~J.; Baddorf,~A.~P.;
	Gupta,~P.; Jain,~H.; Williams,~D.~B.; Gruverman,~A. {Vector Piezoresponse
		Force Microscopy}. \emph{Microscopy and Microanalysis} \textbf{2006},
	\emph{12}, 206--220\relax
	\mciteBstWouldAddEndPuncttrue
	\mciteSetBstMidEndSepPunct{\mcitedefaultmidpunct}
	{\mcitedefaultendpunct}{\mcitedefaultseppunct}\relax
	\EndOfBibitem
	\bibitem[Zhang \latin{et~al.}(2021)Zhang, Chen, Tang, Liao, Di, Mu, Peng, and
	Xiong]{Zhang.2021des}
	Zhang,~H.-Y.; Chen,~X.-G.; Tang,~Y.-Y.; Liao,~W.-Q.; Di,~F.-F.; Mu,~X.;
	Peng,~H.; Xiong,~R.-G. {PFM (piezoresponse force microscopy)-aided design for
		molecular ferroelectrics}. \emph{Chemical Society Reviews} \textbf{2021},
	\emph{50}, 8248--8278\relax
	\mciteBstWouldAddEndPuncttrue
	\mciteSetBstMidEndSepPunct{\mcitedefaultmidpunct}
	{\mcitedefaultendpunct}{\mcitedefaultseppunct}\relax
	\EndOfBibitem
	\bibitem[Gruverman \latin{et~al.}(2019)Gruverman, Alexe, and
	Meier]{Gruverman.2019}
	Gruverman,~A.; Alexe,~M.; Meier,~D. {Piezoresponse force microscopy and
		nanoferroic phenomena}. \emph{Nature Communications} \textbf{2019},
	\emph{10}, 1661\relax
	\mciteBstWouldAddEndPuncttrue
	\mciteSetBstMidEndSepPunct{\mcitedefaultmidpunct}
	{\mcitedefaultendpunct}{\mcitedefaultseppunct}\relax
	\EndOfBibitem
	\bibitem[Kalinin \latin{et~al.}(2007)Kalinin, Rodriguez, Jesse, Karapetian,
	Mirman, Eliseev, and Morozovska]{Kalinin.2007}
	Kalinin,~S.~V.; Rodriguez,~B.~J.; Jesse,~S.; Karapetian,~E.; Mirman,~B.;
	Eliseev,~E.~A.; Morozovska,~A.~N. {Nanoscale Electromechanics of
		Ferroelectric and Biological Systems: A New Dimension in Scanning Probe
		Microscopy}. \emph{Annual Review of Materials Research} \textbf{2007},
	\emph{37}, 189--238\relax
	\mciteBstWouldAddEndPuncttrue
	\mciteSetBstMidEndSepPunct{\mcitedefaultmidpunct}
	{\mcitedefaultendpunct}{\mcitedefaultseppunct}\relax
	\EndOfBibitem
	\bibitem[Li \latin{et~al.}(2014)Li, Wang, Zhang, and Su]{Li.2014}
	Li,~X.; Wang,~B.; Zhang,~T.-Y.; Su,~Y. {Water Adsorption and Dissociation on
		BaTiO$_3$ Single-Crystal Surfaces}. \emph{The Journal of Physical Chemistry
		C} \textbf{2014}, \emph{118}, 15910--15918\relax
	\mciteBstWouldAddEndPuncttrue
	\mciteSetBstMidEndSepPunct{\mcitedefaultmidpunct}
	{\mcitedefaultendpunct}{\mcitedefaultseppunct}\relax
	\EndOfBibitem
	\bibitem[Killgore \latin{et~al.}(2022)Killgore, Robins, and
	Collins]{Killgore.2022}
	Killgore,~J.~P.; Robins,~L.; Collins,~L. {Electrostatically-blind quantitative
		piezoresponse force microscopy free of distributed-force artifacts}.
	\emph{Nanoscale Advances} \textbf{2022}, \emph{4}, 2036--2045\relax
	\mciteBstWouldAddEndPuncttrue
	\mciteSetBstMidEndSepPunct{\mcitedefaultmidpunct}
	{\mcitedefaultendpunct}{\mcitedefaultseppunct}\relax
	\EndOfBibitem
	\bibitem[Weis and Gaylord(1985)Weis, and Gaylord]{Weis.1985}
	Weis,~R.~S.; Gaylord,~T.~K. {Lithium niobate: Summary of physical properties
		and crystal structure}. \emph{Applied Physics A} \textbf{1985}, \emph{37},
	191--203\relax
	\mciteBstWouldAddEndPuncttrue
	\mciteSetBstMidEndSepPunct{\mcitedefaultmidpunct}
	{\mcitedefaultendpunct}{\mcitedefaultseppunct}\relax
	\EndOfBibitem
	\bibitem[Berlincourt and Jaffe(1958)Berlincourt, and Jaffe]{Berlincourt.1958}
	Berlincourt,~D.; Jaffe,~H. {Elastic and Piezoelectric Coefficients of
		Single-Crystal Barium Titanate}. \emph{Physical Review} \textbf{1958},
	\emph{111}, 143--148\relax
	\mciteBstWouldAddEndPuncttrue
	\mciteSetBstMidEndSepPunct{\mcitedefaultmidpunct}
	{\mcitedefaultendpunct}{\mcitedefaultseppunct}\relax
	\EndOfBibitem
	\bibitem[Lindsay \latin{et~al.}(2022)Lindsay, Gaston, Permann, Miller,
	Andr{\v{s}}, Slaughter, Kong, Hansel, Carlsen, Icenhour, Harbour, Giudicelli,
	Stogner, German, Badger, Biswas, Chapuis, Green, Hales, Hu, Jiang, Jung,
	Matthews, Miao, Novak, Peterson, Prince, Rovinelli, Schunert, Schwen,
	Spencer, Veeraraghavan, Recuero, Yushu, Wang, Wilkins, and
	Wong]{lindsay2022moose}
	Lindsay,~A.~D.; Gaston,~D.~R.; Permann,~C.~J.; Miller,~J.~M.; Andr{\v{s}},~D.;
	Slaughter,~A.~E.; Kong,~F.; Hansel,~J.; Carlsen,~R.~W.; Icenhour,~C.;
	Harbour,~L.; Giudicelli,~G.~L.; Stogner,~R.~H.; German,~P.; Badger,~J.;
	Biswas,~S.; Chapuis,~L.; Green,~C.; Hales,~J.; Hu,~T. \latin{et~al.}  2.0 -
	{MOOSE}: Enabling massively parallel multiphysics simulation.
	\emph{{SoftwareX}} \textbf{2022}, \emph{20}, 101202\relax
	\mciteBstWouldAddEndPuncttrue
	\mciteSetBstMidEndSepPunct{\mcitedefaultmidpunct}
	{\mcitedefaultendpunct}{\mcitedefaultseppunct}\relax
	\EndOfBibitem
	\bibitem[Prosandeev and Bellaiche(2007)Prosandeev, and
	Bellaiche]{Prosandeev2007}
	Prosandeev,~S.; Bellaiche,~L. {Asymmetric screening of the depolarizing field
		in a ferroelectric thin film}. \emph{Physical Review B} \textbf{2007},
	\emph{75}, 172109\relax
	\mciteBstWouldAddEndPuncttrue
	\mciteSetBstMidEndSepPunct{\mcitedefaultmidpunct}
	{\mcitedefaultendpunct}{\mcitedefaultseppunct}\relax
	\EndOfBibitem
	\bibitem[Wieder(1955)]{Wieder1955}
	Wieder,~H.~H. {Electrical Behavior of Barium Titanate Single Crystals at Low
		Temperatures}. \emph{Physical Review} \textbf{1955}, \emph{99},
	1161--1165\relax
	\mciteBstWouldAddEndPuncttrue
	\mciteSetBstMidEndSepPunct{\mcitedefaultmidpunct}
	{\mcitedefaultendpunct}{\mcitedefaultseppunct}\relax
	\EndOfBibitem
	\bibitem[Hlinka and Marton(2006)Hlinka, and
	Marton]{Hlinka2006PhenomenologicalMO}
	Hlinka,~J.; Marton,~P. Phenomenological model of a 90$^\circ$ domain wall in
	BaTiO$_3$-type ferroelectrics. \emph{Physical Review B} \textbf{2006},
	\emph{74}, 104104\relax
	\mciteBstWouldAddEndPuncttrue
	\mciteSetBstMidEndSepPunct{\mcitedefaultmidpunct}
	{\mcitedefaultendpunct}{\mcitedefaultseppunct}\relax
	\EndOfBibitem
	\bibitem[Wang \latin{et~al.}(2010)Wang, Meng, Ma, Xu, and Chen]{Wang2010}
	Wang,~J.~J.; Meng,~F.~Y.; Ma,~X.~Q.; Xu,~M.~X.; Chen,~L.~Q. {Lattice, elastic,
		polarization, and electrostrictive properties of BaTiO3 from
		first-principles}. \emph{Journal of Applied Physics} \textbf{2010},
	\emph{108}, 034107\relax
	\mciteBstWouldAddEndPuncttrue
	\mciteSetBstMidEndSepPunct{\mcitedefaultmidpunct}
	{\mcitedefaultendpunct}{\mcitedefaultseppunct}\relax
	\EndOfBibitem
	\bibitem[Mehta \latin{et~al.}(1973)Mehta, Silverman, and Jacobs]{Mehta1973}
	Mehta,~R.~R.; Silverman,~B.~D.; Jacobs,~J.~T. {Depolarization fields in thin
		ferroelectric films}. \emph{Journal of Applied Physics} \textbf{1973},
	\emph{44}, 3379--3385\relax
	\mciteBstWouldAddEndPuncttrue
	\mciteSetBstMidEndSepPunct{\mcitedefaultmidpunct}
	{\mcitedefaultendpunct}{\mcitedefaultseppunct}\relax
	\EndOfBibitem
	\bibitem[Kopal \latin{et~al.}(1997)Kopal, Bahnik, and Fousek]{Kopal1997}
	Kopal,~A.; Bahnik,~T.; Fousek,~J. {Domain formation in thin ferroelectric
		films: The role of depolarization energy}. \emph{Ferroelectrics}
	\textbf{1997}, \emph{202}, 267--274\relax
	\mciteBstWouldAddEndPuncttrue
	\mciteSetBstMidEndSepPunct{\mcitedefaultmidpunct}
	{\mcitedefaultendpunct}{\mcitedefaultseppunct}\relax
	\EndOfBibitem
	\bibitem[Aguado-Puente and Junquera(2008)Aguado-Puente, and
	Junquera]{Aguado-Puente2008}
	Aguado-Puente,~P.; Junquera,~J. {Ferromagneticlike Closure Domains in
		Ferroelectric Ultrathin Films: First-Principles Simulations}. \emph{Physical
		Review Letters} \textbf{2008}, \emph{100}, 177601\relax
	\mciteBstWouldAddEndPuncttrue
	\mciteSetBstMidEndSepPunct{\mcitedefaultmidpunct}
	{\mcitedefaultendpunct}{\mcitedefaultseppunct}\relax
	\EndOfBibitem
	\bibitem[Hong \latin{et~al.}(2017)Hong, Damodaran, Xue, Hsu, Britson, Yadav,
	Nelson, Wang, Scott, Martin, Ramesh, and Chen]{Hong2017}
	Hong,~Z.; Damodaran,~A.~R.; Xue,~F.; Hsu,~S.-L.; Britson,~J.; Yadav,~A.~K.;
	Nelson,~C.~T.; Wang,~J.-J.; Scott,~J.~F.; Martin,~L.~W.; Ramesh,~R.;
	Chen,~L.-Q. {Stability of Polar Vortex Lattice in Ferroelectric
		Superlattices}. \emph{Nano Letters} \textbf{2017}, \emph{17},
	2246--2252\relax
	\mciteBstWouldAddEndPuncttrue
	\mciteSetBstMidEndSepPunct{\mcitedefaultmidpunct}
	{\mcitedefaultendpunct}{\mcitedefaultseppunct}\relax
	\EndOfBibitem
	\bibitem[Wada \latin{et~al.}(2006)Wada, Yako, Yokoo, Kakemoto, and
	Tsurumi]{Wada2006}
	Wada,~S.; Yako,~K.; Yokoo,~K.; Kakemoto,~H.; Tsurumi,~T. {Domain Wall
		Engineering in Barium Titanate Single Crystals for Enhanced Piezoelectric
		Properties}. \emph{Ferroelectrics} \textbf{2006}, \emph{334}, 17--27\relax
	\mciteBstWouldAddEndPuncttrue
	\mciteSetBstMidEndSepPunct{\mcitedefaultmidpunct}
	{\mcitedefaultendpunct}{\mcitedefaultseppunct}\relax
	\EndOfBibitem
	\bibitem[Hlinka \latin{et~al.}(2009)Hlinka, Ondrejkovic, and
	Marton]{Hlinka2009}
	Hlinka,~J.; Ondrejkovic,~P.; Marton,~P. {The piezoelectric response of
		nanotwinned BaTiO$_3$}. \emph{Nanotechnology} \textbf{2009}, \emph{20},
	105709\relax
	\mciteBstWouldAddEndPuncttrue
	\mciteSetBstMidEndSepPunct{\mcitedefaultmidpunct}
	{\mcitedefaultendpunct}{\mcitedefaultseppunct}\relax
	\EndOfBibitem
	\bibitem[Ghosh \latin{et~al.}(2014)Ghosh, Sakata, Carter, Thomas, Han, Nino,
	and Jones]{Ghosh2014}
	Ghosh,~D.; Sakata,~A.; Carter,~J.; Thomas,~P.~A.; Han,~H.; Nino,~J.~C.;
	Jones,~J.~L. {Domain Wall Displacement is the Origin of Superior Permittivity
		and Piezoelectricity in BaTiO$_3$ at Intermediate Grain Sizes}.
	\emph{Advanced Functional Materials} \textbf{2014}, \emph{24}, 885--896\relax
	\mciteBstWouldAddEndPuncttrue
	\mciteSetBstMidEndSepPunct{\mcitedefaultmidpunct}
	{\mcitedefaultendpunct}{\mcitedefaultseppunct}\relax
	\EndOfBibitem
	\bibitem[Pitike \latin{et~al.}(2018)Pitike, Mangeri, Whitelock, Patel, Dyer,
	Alpay, and Nakhmanson]{Pitike.2018}
	Pitike,~K.~C.; Mangeri,~J.; Whitelock,~H.; Patel,~T.; Dyer,~P.; Alpay,~S.~P.;
	Nakhmanson,~S. {Metastable vortex-like polarization textures in ferroelectric
		nanoparticles of different shapes and sizes}. \emph{Journal of Applied
		Physics} \textbf{2018}, \emph{124}, 064104\relax
	\mciteBstWouldAddEndPuncttrue
	\mciteSetBstMidEndSepPunct{\mcitedefaultmidpunct}
	{\mcitedefaultendpunct}{\mcitedefaultseppunct}\relax
	\EndOfBibitem
	\bibitem[Rodriguez \latin{et~al.}(2007)Rodriguez, Callahan, Kalinin, and
	Proksch]{Rodriguez.2007}
	Rodriguez,~B.~J.; Callahan,~C.; Kalinin,~S.~V.; Proksch,~R. {Dual-frequency
		resonance-tracking atomic force microscopy}. \emph{Nanotechnology}
	\textbf{2007}, \emph{18}, 475504\relax
	\mciteBstWouldAddEndPuncttrue
	\mciteSetBstMidEndSepPunct{\mcitedefaultmidpunct}
	{\mcitedefaultendpunct}{\mcitedefaultseppunct}\relax
	\EndOfBibitem
	\bibitem[Karvounis \latin{et~al.}(2020)Karvounis, Timpu, Vogler-Neuling, Savo,
	and Grange]{Karvounis.2020}
	Karvounis,~A.; Timpu,~F.; Vogler-Neuling,~V.~V.; Savo,~R.; Grange,~R. {Barium
		Titanate Nanostructures and Thin Films for Photonics}. \emph{Advanced Optical
		Materials} \textbf{2020}, \emph{8}, 2001249\relax
	\mciteBstWouldAddEndPuncttrue
	\mciteSetBstMidEndSepPunct{\mcitedefaultmidpunct}
	{\mcitedefaultendpunct}{\mcitedefaultseppunct}\relax
	\EndOfBibitem
	\bibitem[Rendón-Barraza \latin{et~al.}(2019)Rendón-Barraza, Timpu, Grange,
	and Brasselet]{Rendón-Barraza.2019}
	Rendón-Barraza,~C.; Timpu,~F.; Grange,~R.; Brasselet,~S. {Crystalline
		heterogeneity in single ferroelectric nanocrystals revealed by polarized
		nonlinear microscopy}. \emph{Scientific Reports} \textbf{2019}, \emph{9},
	1670\relax
	\mciteBstWouldAddEndPuncttrue
	\mciteSetBstMidEndSepPunct{\mcitedefaultmidpunct}
	{\mcitedefaultendpunct}{\mcitedefaultseppunct}\relax
	\EndOfBibitem
	\bibitem[Mahata \latin{et~al.}(2020)Mahata, Koppe, Kumar, Hofsäss, and
	Vetter]{Mahata.2020}
	Mahata,~M.~K.; Koppe,~T.; Kumar,~K.; Hofsäss,~H.; Vetter,~U. {Upconversion
		photoluminescence of Ho$^{3+}$-Yb$^{3+}$ doped barium titanate
		nanocrystallites: Optical tools for structural phase detection and
		temperature probing}. \emph{Scientific Reports} \textbf{2020}, \emph{10},
	8775\relax
	\mciteBstWouldAddEndPuncttrue
	\mciteSetBstMidEndSepPunct{\mcitedefaultmidpunct}
	{\mcitedefaultendpunct}{\mcitedefaultseppunct}\relax
	\EndOfBibitem
	\bibitem[Hao \latin{et~al.}(2011)Hao, Zhang, and Wei]{Hao.2011}
	Hao,~J.; Zhang,~Y.; Wei,~X. {Electric-Induced Enhancement and Modulation of
		Upconversion Photoluminescence in Epitaxial BaTiO$_3$:Yb/Er Thin Films}.
	\emph{Angewandte Chemie International Edition} \textbf{2011}, \emph{50},
	6876--6880\relax
	\mciteBstWouldAddEndPuncttrue
	\mciteSetBstMidEndSepPunct{\mcitedefaultmidpunct}
	{\mcitedefaultendpunct}{\mcitedefaultseppunct}\relax
	\EndOfBibitem
	\bibitem[Sun \latin{et~al.}(2017)Sun, Wang, Pan, Liu, Chen, and Ye]{Sun.2017}
	Sun,~Q.; Wang,~W.; Pan,~Y.; Liu,~Z.; Chen,~X.; Ye,~M. {Effects of temperature
		and electric field on upconversion luminescence in Er$^{3+}$-Yb$^{3+}$
		codoped Ba$_{0.8}$Sr$_{0.2}$TiO$_3$ ferroelectric ceramics}. \emph{Journal of
		the American Ceramic Society} \textbf{2017}, \emph{100}, 4661--4669\relax
	\mciteBstWouldAddEndPuncttrue
	\mciteSetBstMidEndSepPunct{\mcitedefaultmidpunct}
	{\mcitedefaultendpunct}{\mcitedefaultseppunct}\relax
	\EndOfBibitem
	\bibitem[Rodr\'iguez-Carvajal(1993)]{Rodríguez-Carvajal.1993}
	Rodr\'iguez-Carvajal,~J. {Recent advances in magnetic structure determination
		by neutron powder diffraction}. \emph{Physica B: Condensed Matter}
	\textbf{1993}, \emph{192}, 55--69\relax
	\mciteBstWouldAddEndPuncttrue
	\mciteSetBstMidEndSepPunct{\mcitedefaultmidpunct}
	{\mcitedefaultendpunct}{\mcitedefaultseppunct}\relax
	\EndOfBibitem
	\bibitem[Balke \latin{et~al.}(2015)Balke, Maksymovych, Jesse, Herklotz, Tselev,
	Eom, Kravchenko, Yu, and Kalinin]{Balke.2015}
	Balke,~N.; Maksymovych,~P.; Jesse,~S.; Herklotz,~A.; Tselev,~A.; Eom,~C.-B.;
	Kravchenko,~I.~I.; Yu,~P.; Kalinin,~S.~V. {Differentiating Ferroelectric and
		Nonferroelectric Electromechanical Effects with Scanning Probe Microscopy}.
	\emph{ACS Nano} \textbf{2015}, \emph{9}, 6484--6492\relax
	\mciteBstWouldAddEndPuncttrue
	\mciteSetBstMidEndSepPunct{\mcitedefaultmidpunct}
	{\mcitedefaultendpunct}{\mcitedefaultseppunct}\relax
	\EndOfBibitem
	\bibitem[Schindelin \latin{et~al.}(2012)Schindelin, Arganda-Carreras, Frise,
	Kaynig, Longair, Pietzsch, Preibisch, Rueden, Saalfeld, Schmid, Tinevez,
	White, Hartenstein, Eliceiri, Tomancak, and Cardona]{Schindelin.2012}
	Schindelin,~J.; Arganda-Carreras,~I.; Frise,~E.; Kaynig,~V.; Longair,~M.;
	Pietzsch,~T.; Preibisch,~S.; Rueden,~C.; Saalfeld,~S.; Schmid,~B.;
	Tinevez,~J.-Y.; White,~D.~J.; Hartenstein,~V.; Eliceiri,~K.; Tomancak,~P.;
	Cardona,~A. {Fiji: an open-source platform for biological-image analysis}.
	\emph{Nature Methods} \textbf{2012}, \emph{9}, 676--682\relax
	\mciteBstWouldAddEndPuncttrue
	\mciteSetBstMidEndSepPunct{\mcitedefaultmidpunct}
	{\mcitedefaultendpunct}{\mcitedefaultseppunct}\relax
	\EndOfBibitem
\end{mcitethebibliography}
\end{document}